\documentclass[twoside,twocolumn,9pt]{article}
\usepackage{extsizes}
\usepackage[super,sort&compress,comma]{natbib} 
\usepackage[version=3]{mhchem}
\usepackage[left=1.5cm, right=1.5cm, top=1.785cm, bottom=2.0cm]{geometry}
\usepackage{balance}
\usepackage{mathptmx}
\usepackage{mathpazo}
\usepackage{sectsty}
\usepackage{graphicx} 
\usepackage{lastpage}
\usepackage[format=plain,justification=justified,singlelinecheck=false,font={stretch=1.125,small,sf},labelfont=bf,labelsep=space]{caption}
\usepackage{float}
\usepackage{fancyhdr}
\usepackage{fnpos}
\usepackage[english]{babel}
\addto{\captionsenglish}{%
  
}
\usepackage{array}
\usepackage{droidsans}
\usepackage{charter}
\usepackage[T1]{fontenc}
\usepackage[usenames,dvipsnames]{xcolor}
\usepackage{setspace}
\usepackage[compact]{titlesec}
\usepackage{hyperref}

\usepackage{epstopdf}

\definecolor{cream}{RGB}{222,217,201}

\newcommand*\bmu[0]{\mbox{\boldmath $\mu$}}
\newcommand*\balpha[0]{\mbox{\boldmath $\alpha$}}

\newcommand{\comment}[1]{}
\newcommand{\etal}{{\emph{et al. }}}

\begin{document}

\pagestyle{fancy}
\thispagestyle{plain}
\fancypagestyle{plain}{
\renewcommand{\headrulewidth}{0pt}
}

\makeFNbottom
\makeatletter
\renewcommand\LARGE{\@setfontsize\LARGE{15pt}{17}}
\renewcommand\Large{\@setfontsize\Large{12pt}{14}}
\renewcommand\large{\@setfontsize\large{10pt}{12}}
\renewcommand\footnotesize{\@setfontsize\footnotesize{7pt}{10}}
\makeatother

\renewcommand{\thefootnote}{\fnsymbol{footnote}}
\renewcommand\footnoterule{\vspace*{1pt}%
\color{cream}\hrule width 3.5in height 0.4pt \color{black}\vspace*{5pt}} 
\setcounter{secnumdepth}{5}

\makeatletter 
\renewcommand\@biblabel[1]{#1}            
\renewcommand\@makefntext[1]%
{\noindent\makebox[0pt][r]{\@thefnmark\,}#1}
\makeatother 
\renewcommand{\figurename}{\small{Fig.}~}
\sectionfont{\sffamily\Large}
\subsectionfont{\normalsize}
\subsubsectionfont{\bf}
\setstretch{1.125} 
\setlength{\skip\footins}{0.8cm}
\setlength{\footnotesep}{0.25cm}
\setlength{\jot}{10pt}
\titlespacing*{\section}{0pt}{4pt}{4pt}
\titlespacing*{\subsection}{0pt}{15pt}{1pt}

\fancyfoot{}
\fancyfoot[LO,RE]{\vspace{-7.1pt}\includegraphics[height=9pt]{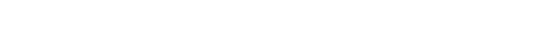}}
\fancyfoot[CO]{\vspace{-7.1pt}\hspace{13.2cm}\includegraphics{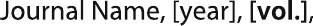}}
\fancyfoot[CE]{\vspace{-7.2pt}\hspace{-14.2cm}\includegraphics{head_foot/RF}}
\fancyfoot[RO]{\footnotesize{\sffamily{1--\pageref{LastPage} ~\textbar  \hspace{2pt}\thepage}}}
\fancyfoot[LE]{\footnotesize{\sffamily{\thepage~\textbar\hspace{3.45cm} 1--\pageref{LastPage}}}}
\fancyhead{}
\renewcommand{\headrulewidth}{0pt} 
\renewcommand{\footrulewidth}{0pt}
\setlength{\arrayrulewidth}{1pt}
\setlength{\columnsep}{6.5mm}
\setlength\bibsep{1pt}

\makeatletter 
\newlength{\figrulesep} 
\setlength{\figrulesep}{0.5\textfloatsep} 

\newcommand{\topfigrule}{\vspace*{-1pt}%
\noindent{\color{cream}\rule[-\figrulesep]{\columnwidth}{1.5pt}} }

\newcommand{\botfigrule}{\vspace*{-2pt}%
\noindent{\color{cream}\rule[\figrulesep]{\columnwidth}{1.5pt}} }

\newcommand{\dblfigrule}{\vspace*{-1pt}%
\noindent{\color{cream}\rule[-\figrulesep]{\textwidth}{1.5pt}} }

\makeatother

\twocolumn[
  \begin{@twocolumnfalse}
{\includegraphics[height=30pt]{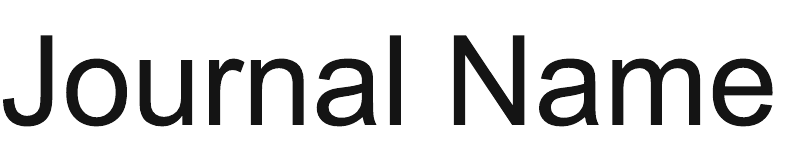}\hfill\raisebox{0pt}[0pt][0pt]{\includegraphics[height=55pt]{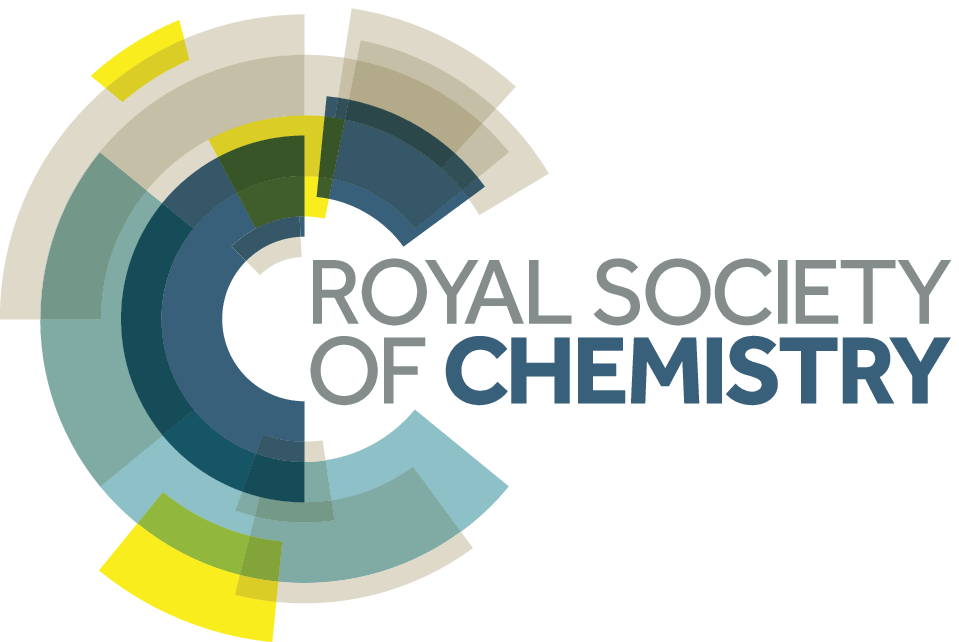}}\\[1ex]
\includegraphics[width=18.5cm]{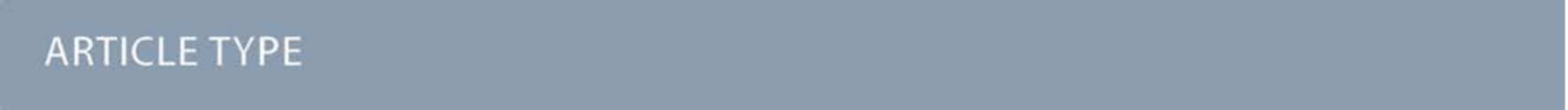}}\par
\vspace{1em}
\sffamily
\begin{tabular}{m{4.5cm} p{13.5cm} }

\includegraphics{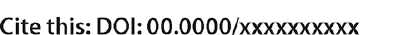} & \noindent\LARGE{\textbf{An atomistic description of alloys and core shells nanoparticles}} \\
\vspace{0.3cm} & \vspace{0.3cm} \\

 & \noindent\large{Lasse K. S{\o}rensen,$^{\ast}$\textit{$^{a}$} Anton D. Utyushev,\textit{$^{b,c}$}  Vadim I. Zakomirnyi,\textit{$^{a,b,c}$}  and Hans \r Agren\textit{$^{a,d,e}$}} \\

\includegraphics{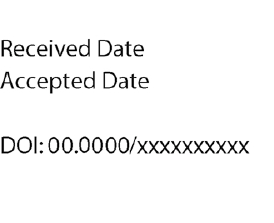} & \noindent\normalsize{Using the extended discrete interaction model we investigate the tuneabilty of surface plasmon resonance in alloys and core-shell nanoparticles made from silver and gold. We show that the surface plasmon resonance of these alloys and core-shell particles to a large extent follow Vegard's law irrespective of the geometry of the nanoparticle. We show the evolution of the polarizability with size and demonstrate the highly non-linear behaviour of the polarizability with the ratio of the constituents and geometry in alloys and core-shell nanoparticles, with the exception for nanorod alloys. A thorough statistical investigation reveals that there is only a small dependence of the surface plasmon resonance on atomic arrangement and exact distribution in a nanoparticle and that the standard deviation decrease rapidly with the size of the nanoparticles. The physical reasoning for the random distribution algorithm for alloys in discrete interaction models is explained in details and verified by the statistical analysis.  
} \\

\end{tabular}

 \end{@twocolumnfalse} \vspace{0.6cm}

  ]

\renewcommand*\rmdefault{bch}\normalfont\upshape
\rmfamily
\section*{}
\vspace{-1cm}


\footnotetext{\textit{$^{a}$~Department of Theoretical Chemistry and Biology, School of Engineering Sciences in Chemistry, Biotechnology and Health, Royal Institute of Technology, Stockholm, SE-10691, Sweden.}}
\footnotetext{\textit{$^{b}$~Siberian Federal University, Krasnoyarsk, 660041, Russia.}}
\footnotetext{\textit{$^{c}$~Institute of Computational Modeling, Federal Research Center KSC SB RAS, Krasnoyarsk, 660036, Russia.}}
\footnotetext{\textit{$^{d}$~Federal Siberian Research Clinical Centre under FMBA of Russia, 660037, Kolomenskaya, 26  Krasnoyarsk, Russia.}}
\footnotetext{\textit{$^{e}$~College of Chemistry and Chemical Engineering, Henan University, Kaifeng, Henan 475004 P. R. China.}}

\footnotetext{$^{\ast}$~Corresponding author: lasse.kragh.soerensen@gmail.com}




\section{\label{sec:intro} Introduction}

Like for plasmonic nanoparticles in general, there is a great deal of interest in bimetallic, or alloyed, small nanoparticles~\cite{Sharma2019, Loza2020} due to their potential applications in a number of technological areas,  like bioimaging\cite{Sun2020}, biomedical plasmonic based sensors and devices~\cite{Zakomirnyi2016},  and in  heterogeneous catalysis.  For instance,  the position of plasmon resonances can be adjusted over a wide range of wavelengths by varying the composition of the constituent metals, making bimetallic particles an interesting proposition for use in surface plasmon enhanced imaging. The recent advancement  in their synthesis and characterization~\cite{Wang2016,Yasuhara2020} have thus opened up possibilities to produce such alloyed or bimetallic nanoparticles  for particular purposes and applications. This goes especially for bimetallic particles formed by  noble metal elements like  platinum, gold , silver and copper. Like for the corresponding monometallic plasmonic particles there are requirements on their design with respect to  crystallographic  structure, shape and size and -  for certain applications-  surface functionalization. In addition to what is required  for  monometallic particles, there is also need to characterize the element composition and the internal elementary distribution of the particles and the homogeneity of the particle population. Making alloys or core-shell structures is really the only way to get a decent blue-shift in plasmon resonance frequency since geometry alterations always gives intense red-shift.
As in other areas of nanothechnology  the synthesis and characterization procedures can be greatly boosted by design strategies based on theoretical modelling~\cite{Rasskazov2019,Rasskazov2020}. The bimetallic nature and the new parametric dimensions that appear for such particles with respect to the monometallic ones, pose special requests not easily met by traditional classical plasmonic models. This goes especially for “small” bimetallic nanoparticles for which  the  use of dielectric constants of bulk materials, makes it impossible to take into account structural differentiation for the dielectric response. In particular when the small particles are mixed with different elements and with large surface to volume ratios and thus when the mean free path of the conduction electrons need to be considered.  Here approaches  are called for that are more precisely can relate to the discrete atomic structure of the nanoparticles to the dielectric and plasmonic properties. Unfortunately pure quantum approaches are still only applicable for the very small particles, leaving a size region $1- 15$~nm unattainable by either classical and quantum theory.
Discrete interaction models  have the inherent capacity to deal with size effects down to the atomic level. In a recent work we presented an extended discrete interaction model (ex-DIM) to simulate the geometric dependence of plasmons in the size range of $1-15$~nm where the Clausius-Mossotti relation is replaced by a  static atomic polarizability to obtain the frequency-dependent dielectric function.\cite{exdim}  The static atomic polariziability  was modeled  as the sum of three size-dependent Lorentzian oscillators and, with Gaussian charge distributions and atomic radii that vary with the coordination number. The frequency-dependent Lorentzian oscillators depend on the plasmon length along the three Cartesian directions using the concept of the plasmon length as defined in the work of Ringe \etal.\cite{ringe2012} In this way we extended previous discrete interaction models  to  make possible description of the polarizability of nanoparticles with different size , shape  and composition, and take account of the dependence of the polarizability of the surface topology or structure of the metallic nanoparticles. This extension of the DIM models serves as a necessary step in order to model bimetallic particles.
The purpose of the present work is to use the extended DIM model to explore the appearance of plasmonic excitation in alloyed nanoparticles, taking alloys between gold and silver as example, and explore   how their plasmonic properties evolve with respect to concentration of the two components, how size and shape of the full particles modify the properties,  how these features can be related to the corresponding properties of the  monometallic particles containing either element and to compare with experimental findings that now are available. 


\section{\label{sec:theory} Theory and approximations}

\subsection{ex-DIM}

The extended Discrete Interaction Model (ex-DIM)~\cite{exdim,Zakomirnyi2020}, is a further development of the DIM model~\cite{Jensen2008,jensen_dim} where significant improvements in the description of the surface topology, geometric dependence and parameterization of the SPR(s) are introduced. 
The ex-DIM model has a special applicable edge for systems in the $1-15$~nm size range, where quantum mechanical models cannot be applied, due to the scaling of these methods, and where the concept of a bulk dielectric constant used in classical models breaks down. The $1-15$~nm size range is, however, a very important region since this is the size region where the onset of the SPR(s) is seen and where the nanoparticles are still small enough to be used in bio-medical applications.
Due to the inability of extrapolating data from quantum mechanical methods and classical methods into the $1-15$~nm size range the ex-DIM must be parameterized directly from experimental data.\cite{exdim}
We will in Sec. \ref{sec:sense_struc} briefly show how a new element easily can be parameterized for the ex-DIM and in Sec. \ref{sec:alloy_imp} describe how alloys with a given initial distribution is implemented.

The ex-DIM model is a discrete structure model where each atom is represented by a Gaussian charge distribution and endowed with a polarizability and a capacitance which govern the inter atomic interaction. The Lagrangian is written in the usual form as the interaction energy $E$ minus the charge equilibration constraint expressed via the Lagrangian multiplier $\lambda$:
\begin{equation}
\label{tm:te}
L[\{\bmu, q \}, \lambda] = E[\{\bmu, q \}] - \lambda ( q^{tot}-\sum_{i}^{N} q_i) \ ,
\end{equation}
\noindent
where $N$ is the number of atoms, $q_i$ is the fluctuating charge assigned to the $i$-th atom, and $q^{tot}$ is the total charge of the nanoparticle. The interaction energy $E[\{\bmu, q \}]$ in this way captures all different types of interactions involving fluctuating dipoles $\bmu$, charges $q$ and an external field and described in greater detail in ref. \cite{exdim}.

The surface topology is captured by a coordination number, as defined by Grimme~\cite{Grimme2010}, and is assigned to each atom. 
The coordination number $f_{cn}$ modifies the atomic polarizability through the scaling of the radius
 \begin{equation}
 \label{pol}
    \balpha_{ii,kl}(\omega) = \left(\frac{R_i (f_{cn})}{R_{i,{\rm bulk}}}\right)^3 \balpha_{i,s,kl} L(\omega,\mathbf{P})
 \end{equation}
and likewise for the capacitance
\begin{equation}
\label{eq.c_ii_tenzor}
 c_{ii,kl} = \delta_{kl} f_{c}  \ {\rm with}  \ f_{c} =  c_{i,s} \left [ 1 +  d  \frac{R_i(f_{cn})} {R_i(12)}\right] L(\omega,\mathbf{P}). 
\end{equation}
In Eqs. \eqref{pol} and \eqref{eq.c_ii_tenzor}, 
$R_{i,{\rm bulk}}$ is the bulk radius of the atom, 
$R_i(f_{cn})$ the coordination number scaled radius~\cite{exdim}, 
$\balpha_{i,s,kl}$ the static atomic polarizability~\cite{Schwerdtfeger_pol}, 
$d=0.1$ a scaling factor and $L(\omega,\mathbf{P})$ a size-dependent Lorentzian. 
The polarizability and capacitance of alloys will in this way not only have a spacial dependence from the discrete structure from the interaction but also a small one from the modified surface atoms. 

The geometric dependence of the SPR is determined by the size-dependent Lorentzian $L(\omega,\mathbf{P})$
\begin{equation}
    \label{lall}
    L(\omega,\mathbf{P}) = N (L_x(\omega,P_x) + L_y(\omega,P_y) + L_z(\omega,P_z)) ,
\end{equation}
where each Lorentzian depends on the plasmon length $P_i$~\cite{ringe2012} in the given direction
\begin{equation}
 \label{omega_size}
    \omega_{i}(P_i)  = \omega_a (1+ A/P_i ) ,
\end{equation}
and in this way cluster size dependence and complicated geometrical shapes, with up to at least three SPRs, can be simulated for solid particles. Since the atomic radius for different atoms is slightly different the plasmon length $P_i$ is not a constant even when the same discrete structure is used for alloys though the change is only in the difference between the atomic radii of the constituents.
$\omega_a$ and $A$ are the only fitted parameters in the ex-DIM model~\cite{exdim}.

The isotropic polarizability is determined from the fluctuating charges, $q$, and dipoles, $\bmu$ which are determined by inversion of the relay matrix.~\cite{Birge1980a,Thole1981,Jensen2000a,Jensen2000,Applequist1996,Sundberg1978,exdim} In this way all SPRs are presented together in the same spectrum.

\section{Geometry}

A fundamental problem of the geometric models is the need for a discrete structure in the simulations. In molecular physics the geometry of known molecules is usually tabulated and can be read in directly. This, however, is not the case for metallic nanoparticles where neither the discrete structure or even the number of particles are known. Only the overall geometric shape of the cluster with dimension on the nm length scale with some error bars are known and these provide no information about the internal discrete structure. Geometry optimization is also not helpful since geometry optimization is NP-hard and therefore not feasible for clusters with thousands of atoms.\cite{cai_2005,HUANG2011199,Fan201664} We will therefore here briefly discuss the influence of discrete structure and a pragmatic yet accurate approach in which a geometry is easily generated for both pure metals and for alloys.

\subsection{Sensitivity to structure}
\label{sec:sense_struc}

From systematic investigations on small silver and gold nano particles it is evident that there is a clear trend in the evolution of the surface plasmon resonance with size on the nanometer scale.\cite{Harbich1993,Fedrigo1993,Scholl2012,raza2103,nilius2000,charle,lunskens2015,campos2019,bastus2011,piella2016,link1999,paul2018,njoki2007,hong2013} Yet for any given particle size  all measured particles vary in both the discrete structure and particle number. Since the discrete structure differences does not lead to an extreme broadening of the plasmon peak for larger clusters the discrete structure differences instead show that having the exact discrete structure, like in molecular physics, is not of extreme importance for the SPR.  This implies that having the same discrete lattice structure, for which the method has been optimized with, should suffice for all sizes.

\begin{figure}
    \centering
    \includegraphics[width = 0.48\textwidth]{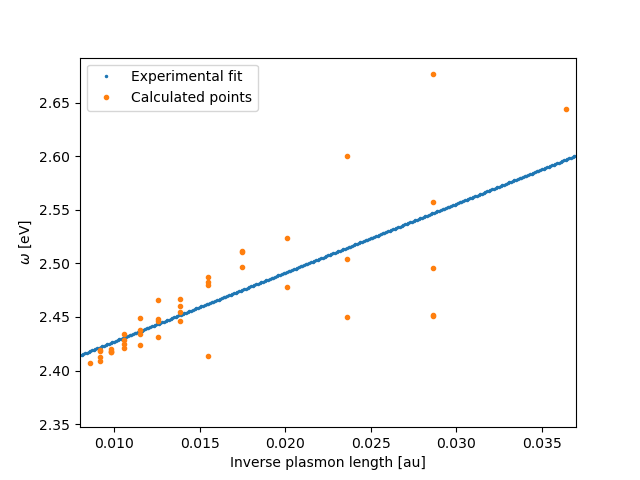}
    \caption{Linear fit of the experimental data\cite{bastus2011,piella2016,link1999} compared with the recalculated clusters using the fitted parameters. The number of atoms in the clusters here vary from 135-7419 with sizes from 1.45-6.18 nm.}
    \label{fig:aufit}
\end{figure}

For smaller structures the sensitivity to the discrete structure is expected to be larger which is also seen in the parameterization of gold in Fig. \ref{fig:aufit}.
This greater variation in the SPR is due to the greater percentage variation in the number of particles for a fixed plasmon length and that with fewer atoms comes an increasing sensitivity to the placement of the individual atoms in the discrete structure.
For clusters with a plasmon length of 1.85 nm we, in this case, see the greatest variation in the surface plasmon resonance, from 2.45-2.67 eV, but we here also have the greatest percentage variation in the particle number since the number of atoms range from 141-249. For larger clusters we see a much smaller variation in the surface plasmon resonance. For example we see that particles with a plasmon length of 5.92 nm the surface plasmon resonance only varies from 2.41-2.42 eV and even though the particle number ranges from 6051 to 7011. 

The parameterization of gold data taken from experiments of gold clusters in solution in the 3.6-17.6 nm size region have been used.\cite{bastus2011,piella2016,link1999}
The parameterization was performed by fitting the experimental data to the inverse plasmon length and afterwards finding an optimum frequency for a set of different clusters. From this $\omega_a$ and $A$ of Eq. \ref{omega_size} could be fitted in the same way that the silver parameters was obtained.\cite{exdim} We here find $\omega_a = 0.0682549$ and $A=6.43041$ in atomic units. Furthermore for the static polarizability we use $\alpha = 36$ au\cite{Schwerdtfeger_pol} and set the broadening $\gamma = 0.0016$ along with the surface and bulk radii of $r_1 = 1.74$ and $r_2=1.56$, respectively.

While the ex-DIM is parameterized from spherical clusters within a $1-15$~nm size range and a limited frequency interval it still  remains valid in a much broader frequency range as shown by calculations on nanorods and nanocubes~\cite{exdim}.
The large frequency range of the ex-DIM  model is possible because any red or blue shift due to geometric distortions from a sphere can be described by the interaction between the atoms in this model and no external data is required for this~\cite{exdim}.
The ex-DIM model is therefor not limited in the frequency range by the parameterization range unlike classical models which are limited by the experimental range for which the dielectric constant have been measured.

\subsection{Geometry of alloy and core-shell clusters}

Like for the pure metals the discrete structure of alloys is not possible to obtain from experiment and contain an added complication since the unit cells of the metals in the alloy will differ. The latter problem of not knowing the lattice parameters of the alloy can usually be overcome using the empirical law from Vegard
\begin{equation}
\label{vegards_law}
a_{A_{1-x}B_x} = (1-x)a_A + x a_B 
\end{equation}
where the alloy lattice parameter $a_{A_{1-x}B_x}$ is approximated by a weighted mean of the two constituents $A,B$ lattice parameters $a_A$ and $a_B$,respectively.\cite{vegard,vegards_law} 

For core shell structures connecting the lattice of the core with the shell is not a simple problem for models using a discrete structure and there does not seem to be a simple way to connect two perfect lattices with different lattice parameters without having to distort these lattices at the boundaries. In order to overcome this we have optimized gold and silver using the same lattice constants which is only possible due to the very little difference in the lattice constants for these two metals. We see the usage of the same lattice constants as a pragmatic approach for this particular type of core-shell structures and not a general solution.

\subsection{Alloy implementation}
\label{sec:alloy_imp}
\begin{figure}
    \centering
    \includegraphics[width = 0.4\textwidth]{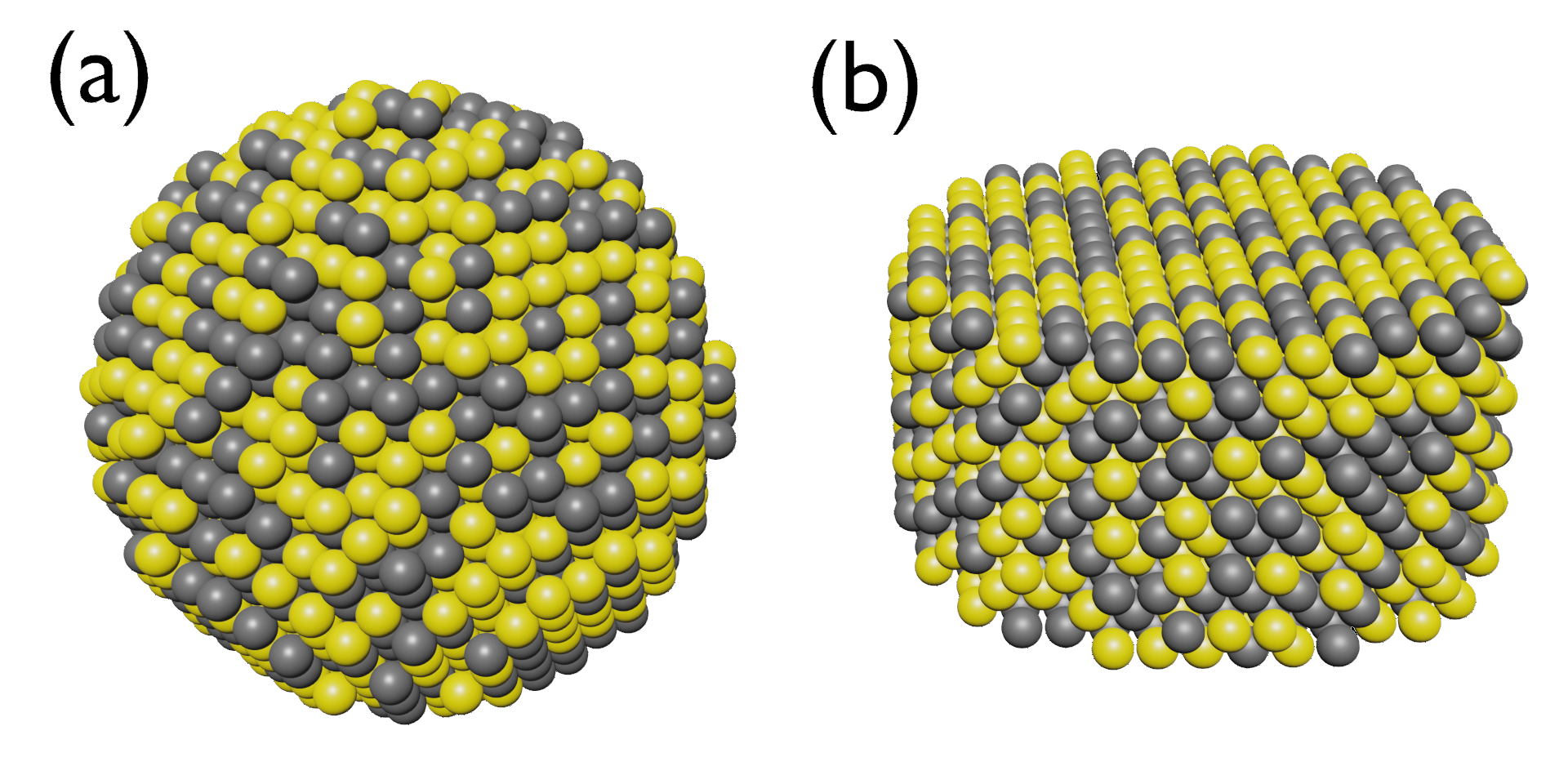}
    \caption{Schematic representation of alloy sphere (a) and disc (b) nanoparticles taken from a random distribution of Au (yellow) and Ag (grey) atoms.}
    \label{fig:sphere_and_disc}
\end{figure}

Since the exact placement of the constituents in an alloy cannot be predicted and differ from cluster to cluster we have chosen to represent this using a simple random selection of elements based on an initial probability distribution. This means that there will be both a random spacial distribution of elements along with a small variation in the ratio between the elements due to the randomized drawing of the elements. In this way no two clusters will be exactly alike and a statistical analysis of the slightly expected broadening of the SPR for alloys can be analyzed in terms of the variation in the ratio and spacial distribution of the constituents and error bars for the alloys can be assessed. Two random distributions for a sphere and disc structure are shown in Fig. \ref{fig:sphere_and_disc}.

\section{\label{sec:results} Results and Discussion}

Even though Vegard's law initially was formulated in order to estimate the lattice parameters of alloys it has often been extended include all properties of alloys
\begin{equation}
\label{general_vegard}
P_{A_{1-x}B_x} = (1-x)P_A + x P_B
\end{equation}
where $P$ is any property.\cite{DUAN20091}
We are here particularly interested in examining if Vegard's law also holds true for the progression of the position of the SPR with mixing of the constituents as indicated in other studies\cite{Pena-Rodriguez:14,skidanenko,link1999,Kuladeep:s,Hidayah_2019,C5TB00644A,C4CP04673K,doi:10.1021/acs.jpcc.0c02630,Tian:18,Zhang:20,Ni:19,doi:10.1021/acs.jpcc.0c02630} for alloys and to some extent core shells or if this is geometry dependent as experiments on nano discs predict.\cite{Nishijima:12}
Since the polarizability or extinction cross section differs for Au and Ag it would also be of interest to see if the strength of the response to an external field also follow Vegard's law since the strength of the response determines strength of the local electric field from the SPR. Here previous results shows a non-linear dependence for the response for spherical clusters.\cite{Pena-Rodriguez:14,skidanenko,link1999} Before this we will perform a statistical analysis to determine how the error varies with spacial and constituent distribution along with size of the cluster for a spherical alloy nanoparticle in order to get an estimation of the error bars when performing a single calculation.

\subsection{Statistical analysis of model}
\label{sec:stat}

The aim of the statistical analysis is to examine the effect of randomly drawing elements and give estimations for error bars since it will not always be possible to make a full statistical analysis for every cluster. Secondary goals are to ensure that the model behaves as would be expected from a random drawing of a finite number of elements with a given probability, to analyse how large the error is for classical models using a simple linear combination of the constituents along with how the standard deviation varies with the size of the cluster.

From the fitting of Au in Fig \ref{fig:aufit} we would expect that the standard deviation of an alloy would decrease with size since the importance of the individual placement of an atom would decrease and less sensitivity to variations in the actual distribution of the constituents should be seen. Secondly from a statistical perspective we would expect that the standard deviation for alloys where one of the constituents is very dominating is lower than one where the distribution is more even.

In Table \ref{tbl:stat} we present a statistical analysis for a set of clusters ranging from 249 to 6051 atoms or 1.85 to 5.79 nm in size with three different distributions in order to show both the size and distribution dependence of the standard deviation of the SPR and extinction cross section. Due to computational resources we reduce the number of sampling points with the size and width of the sample though significant changes with increasing number of sampling points is not expected and all trends are easy to see. Due to the larger standard deviation in the SPR seen for smaller samples the frequency range sampled in Table \ref{tbl:stat} is not the same though the interval between each frequency was constant and set at 0.0001 au. Since the CPU time set for each size, above 1553 atoms, is the same the number of samples for each distribution varies. We have in Table \ref{tbl:stat} added extra digits on both the standard deviation $\sigma$ and the averages $\mu$ in order to better illustrate the trends.

\begin{table*}
\small
  \caption{\ Statistical data from the sampling of Ag and Au alloys of different sizes and distributions showing the average, standard deviation, minimum and maximum values found for the number of Ag atoms in a cluster, the SPR and Extinction cross section per atom. Extra digits have in some cases been added in order to better illustrate trends.}
  \label{tbl:stat}
  \begin{tabular*}{\textwidth}{@{\extracolsep{\fill}}rcc |c| rrrr | cccc| cccc}
    \hline
    \multicolumn{3}{c|}{Distribution} & Samples & \multicolumn{4}{c|}{Ag} & \multicolumn{4}{c|}{SPR [nm]} & \multicolumn{4}{c}{Extinction cross section [nm$^2$/atom]} \\ 
     atoms & \% Ag & \% Au & \# &  $\mu$ & $\sigma$ & min & max & $\mu$ & $\sigma$ & min & max & $\mu$ & $\sigma$ & min & max \\
    \hline
     & 90 & 10 & 9099 & 224 & 5 & 205 & 242 & 348.4 & 1.92 & 342.6 & 356.5 & 0.2486 & 0.0109 & 0.1914 & 0.2866 \\
     249 & 50 & 50 & 9099 & 124.4 & 7.9 & 96 & 157 & 392.9 & 7.85 & 368.9 & 424.6 & 0.1678 & 0.0091 & 0.1331 & 0.1994 \\
     & 30 & 70 & 9099 & 74.7 & 7.2 & 49 & 101 & 431.9 & 10.3 & 402.5 & 460.2 & 0.1496 & 0.0064 & 0.1263 & 0.1773 \\ \hline
     & 90 & 10 & 8810 & 607 & 8 & 578 & 639 & 350.8 & 11.2 & 332.4 & 367.1 & 0.2149 & 0.0075 & 0.1956 & 0.2368 \\
     675 & 50 & 50 & 9099 & 338 & 13 & 290 & 385 & 410.7 & 4.91 & 395.9 & 427.8 & 0.1542 & 0.0038 & 0.1408 & 0.1697 \\
     & 30 & 70 & 9099 & 203 & 12 & 159 & 254 & 448.2 & 4.26 & 430.3 & 462.5 & 0.1416 & 0.0028 & 0.1305 & 0.1541 \\ \hline
     & 90 & 10 & 8134 & 1398 & 12 & 1355 & 1436  & 359.7 & 1.25 & 356.0 & 364.5 & 0.2405 & 0.0030 & 0.2288 & 0.2510 \\
    1553 & 50 & 50 & 5564 & 777 & 20 & 709 & 852 & 412.3 & 2.99 & 405.0 & 421.9 & 0.1620 & 0.0028 & 0.1520 & 0.1725 \\
     & 30 & 70 & 6194 & 465 & 18 & 403 & 535 & 450.6 & 2.79 & 440.2 & 459.2 & 0.1444 & 0.0019 & 0.1363 & 0.1511 \\ \hline
     & 90 & 10 & 2732 & 3030 & 17 & 2976 & 3090  & 367.3 & 0.96 & 363.6 & 370.4 & 0.2368 & 0.0020 & 0.2297 & 0.2436 \\
    3367 & 50 & 50 & 1634 & 1684 & 28 & 1598 & 1784 & 423.1 & 1.71 & 417.2 & 429.0 & 0.1673 & 0.0018 & 0.1622 & 0.1723 \\
     & 30 & 70 & 2621 & 1009 & 26 & 919 & 1121 & 457.5 & 1.60 & 452.0 & 462.6 & 0.1529 & 0.0011 & 0.1488 & 0.1567 \\ \hline
     & 90 & 10 & 1159 & 5446 & 23 & 5372 & 5515 & 370.6 & 0.49 & 369.0 & 372.2 & 0.2575 & 0.0017 & 0.2515 & 0.2624\\
    6051 & 50 & 50 & 746 & 3024 & 38 & 2900 & 3145 & 424.2 & 1.30 & 419.6 & 428.2 & 0.1719 & 0.0013 & 0.1682 & 0.1762 \\
     & 30 & 70 & 809 & 1818 & 33 & 1710 & 1921 & 459.4 & 1.20 & 456.2 & 464.0 & 0.1546 & 0.0009 & 0.1516 & 0.1575 \\
    \hline
  \end{tabular*}
\end{table*}

As expected we see that the standard deviation in the number of Ag atoms increases with both size and evenness in the distribution. For the SPR we also see, as expected from Fig \ref{fig:aufit}, that the standard deviation for the SPR decreases with size and increase with evenness in the distribution. At around 3000 atoms the standard deviation is below 2 nm and therefore below the accuracy which can be expected from any calculation. From an atomistic perspective we therefore also see that the classical way of treating an alloy without any resolution at the atomic level does not introduce any significant error for the SPR of larger systems calculated using classical methods.\cite{silver_dielectric_2014,doi:10.1021/acs.jpcc.0c02630,Kuladeep:s} Furthermore we see that with good size correction the classical methods for alloys can safely be extended to small systems down to 4-5 nm. For alloys where on of the constituents is very dominating, above 90 percent, even smaller systems can be safely simulated. There is, however, one outlier in the data in Table \ref{tbl:stat} namely the 675 atoms cluster with distribution of $90\%$ Ag and $10\%$ Au where the standard deviation for the SPR is 11.2nm where a standard deviation of around 1.5nm would be expected from the trend. This outlier appears because many of the spectra for this 675 atoms cluster shows a double peak, which the maximum alternates between, thereby causing a very broad distribution and hence large standard deviation.

\begin{figure}
    \centering
    \includegraphics[width = 0.48\textwidth]{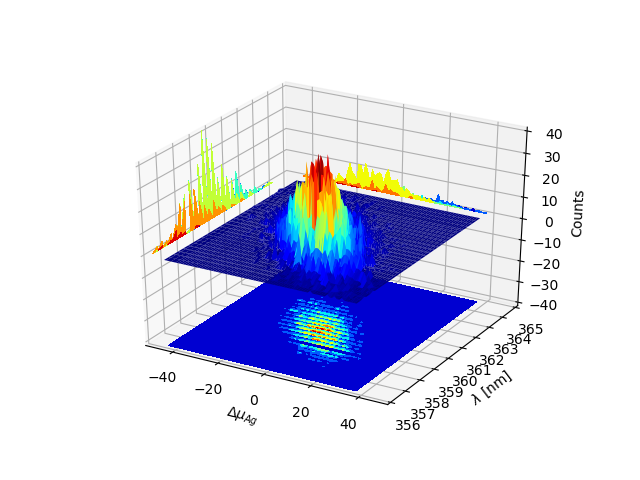}
    \caption{Statistics data for a 1553 atom alloy cluster with a random distribution with a probability distribution of 10 Au 90 Ag.}
    \label{fig:1553_10_90}
\end{figure}

\begin{figure}
    \centering
    \includegraphics[width = 0.48\textwidth]{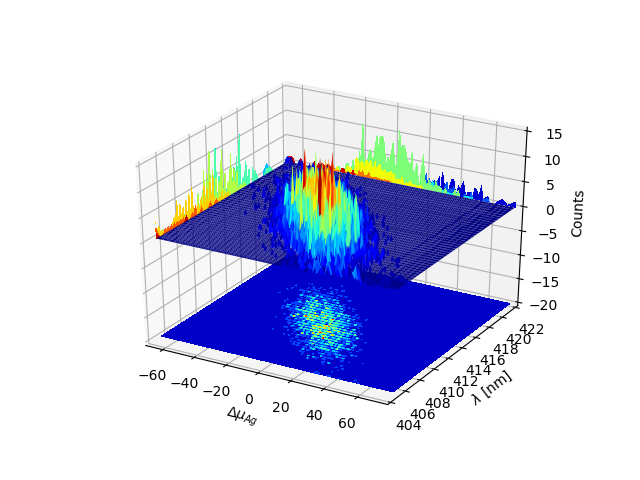}
    \caption{Statistics data for a 1553 atom alloy cluster with a random distribution with a probability distribution of 50 Au 50 Ag.}
    \label{fig:1553_50_50}
\end{figure}

\begin{figure}
    \centering
    \includegraphics[width = 0.48\textwidth]{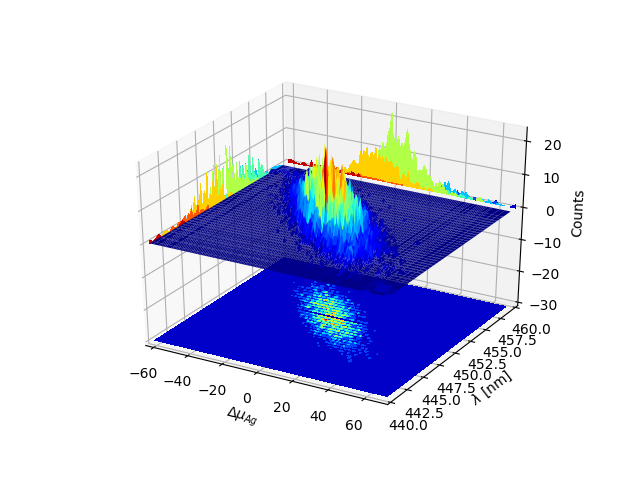}
    \caption{Statistics data for a 1553 atom alloy cluster with a random distribution with a probability distribution of 70 Au 30 Ag.}
    \label{fig:1553_70_30}
\end{figure}

In order to analyse and illustrate the effect of the random distributing on a fixed number of constituents and the variation of the constituents due to the random drawing of the constituents we have chosen to focus on the 1553 atoms cluster with different distributions. The 1553 atoms cluster shows a clear difference in the standard deviation of both the number of Ag atoms and the SPR for the different distribution and a sufficient number of samples could easily be collected as seen in Table \ref{tbl:stat}. In Figs. \ref{fig:1553_10_90}-\ref{fig:1553_70_30} the three different distributions from Table \ref{tbl:stat} of the 1553 atoms cluster is plotted. 

In Fig. \ref{fig:1553_10_90} we clearly see that the SPR follows the change away from the average number of Ag atoms $\Delta \mu_{Ag}$ with a red shift for a negative shift of $\Delta \mu_{Ag}$ and blue shift for a positive shift of $\Delta \mu_{Ag}$ as would be expected. For the clusters with the same number of constituents we also see that the spacial distribution of the atoms also matters though less than the variation in $\Delta \mu_{Ag}$ since we have a very nice and centered peak. 

Comparing the three distributions in Figs. \ref{fig:1553_10_90}-\ref{fig:1553_70_30} we clearly see that the Ag 90 Au 10 distribution in Fig . \ref{fig:1553_10_90} is significantly more narrow and peaked both along $\Delta \mu_{Ag}$ and $\lambda$. This can also be seen from the projection onto the Counts axis where it is clear that the less even the distribution is the more clear and narrow the peak becomes. The significantly smaller standard deviation for the SPR of the Ag 90 Au 10 distribution compared to the Ag 50 Au 50 and Ag 30 Au 70 distributions is also clearly visible from the projection of the distribution onto the $\lambda$ axis.

The extinction cross section per atom in Table \ref{tbl:stat} shows no real variance with size but only with distribution where the extinction cross section increases with the amount of Ag in the alloy. Though the difference in the extinction cross section between Ag 30 Au 70 and Ag 50 Au 50 distributions is significantly smaller than could be expected from Vegard's law. We will analyse this observation in more detail in Sec. \ref{sec:polarizability} when we look at the evolution of the polarizability as a function of the constituents. We here also see the expected trend where the standard deviation decreases with size though remain rather small throughout. We here note that since the broadening of the Lorentzians in Eq. \ref{lall} is not fitted only the trend and not the absolute values of the extinction cross section is to be interpreted here for the ex-DIM.

\subsection{Surface plasmon resonance for alloys and core shells}

Since Vegard's law assumes linearity hence a linear energy scale for the SPR should be used 
\begin{equation}
\label{vegard}
\lambda^{Vegard}(x,R) = (1-x)\lambda_{Au}(R) + x\lambda_{Ag}(R).
\end{equation}
We will in the following use eV and not nm, even if nm is the prevalent choice of unit, since only the former of the two is linear energy unit. 

We will here examine if any non-linearity effects in the SPR can be induced by the geometric structure of the alloy clusters as observed experimentally by Nishijima \etal on nano discs. The experimentally observed a very large red shift in the spectra they reproduced in theoretical predictions using FDTD where the plasma frequency $\omega_p$ and relaxation time $\tau$ was extracted by inserting the experimental data into the Drude model.\cite{Nishijima:12}
We will here examine two spherical clusters with different radii, a nanorod and a nano discs though with different relative dimensions as the ones used experimentally by Nishijima \etal. 
Another possibility of mixing metals is by making core-shell structures. For the core-shell structures we will also examine two geometrical structures namely spheres and rods where we show both Au core and Ag shell along with Ag core and Au shell in order to examine if there is significant difference in which metal is the core and shell. 
Since the atoms in ex-DIM are discrete the steps in the distribution of the metals for core shell structures cannot be divided into equal steps as it can for alloys. Furthermore the use of spherical structures having a percentage wise large core will result in an atomically thin surface and the very large core should therefore be interpreted with care. In appendix \ref{app:a} we plot all spectra from which data have been extracted.

\begin{figure}
    \centering
    \includegraphics[width = 0.48\textwidth]{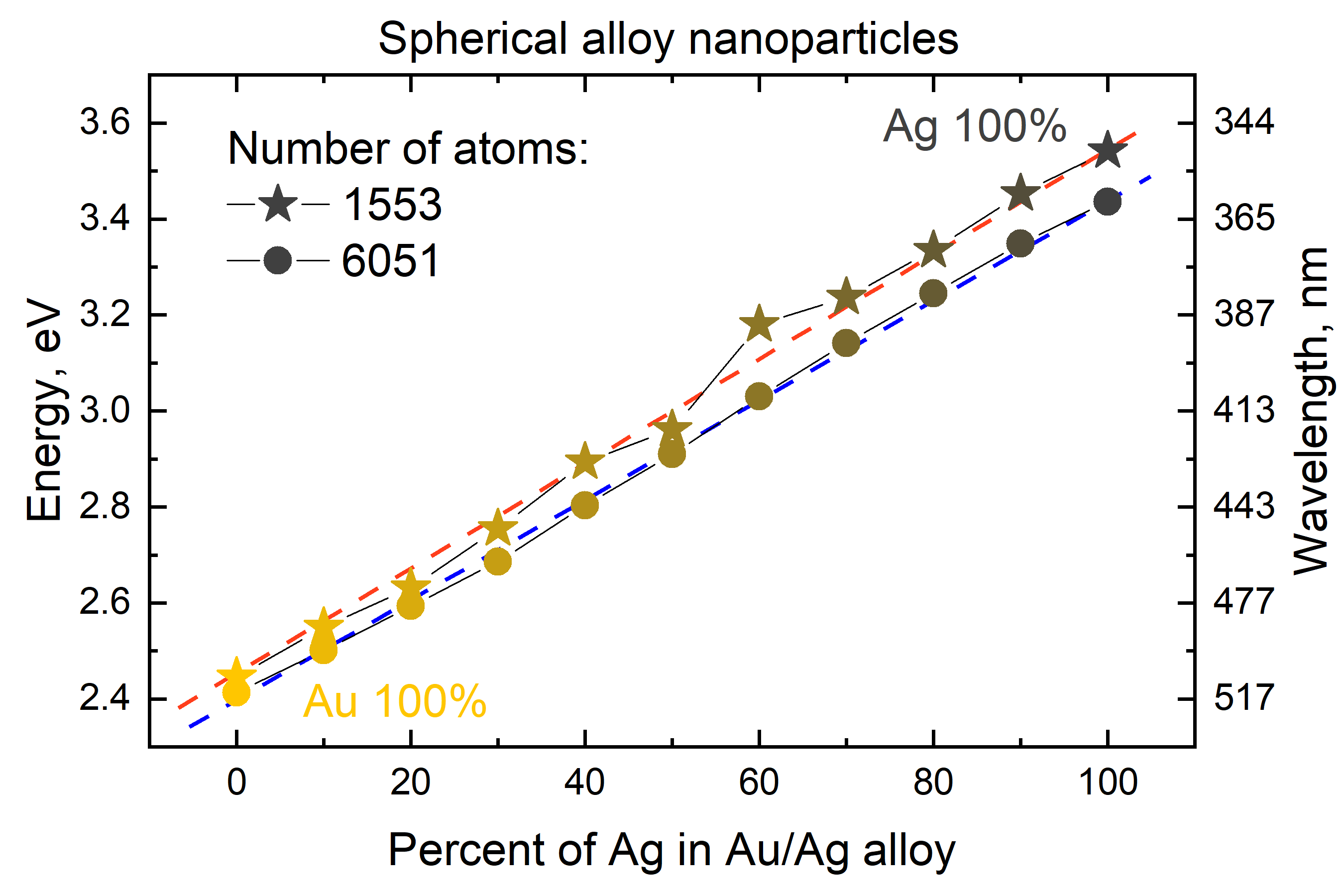}
    \caption{Position of SPR for spherical Au/Ag alloy nanoparticles from 1553 atoms (stars) and 6051 atoms (circles) with different percentage distribution of Au and Ag along with dashed lines showing Vegard's law.}
    \label{fig:fit_plasmon_sphere_1553}
\end{figure}

\begin{figure}
    \centering
    \includegraphics[width = 0.48\textwidth]{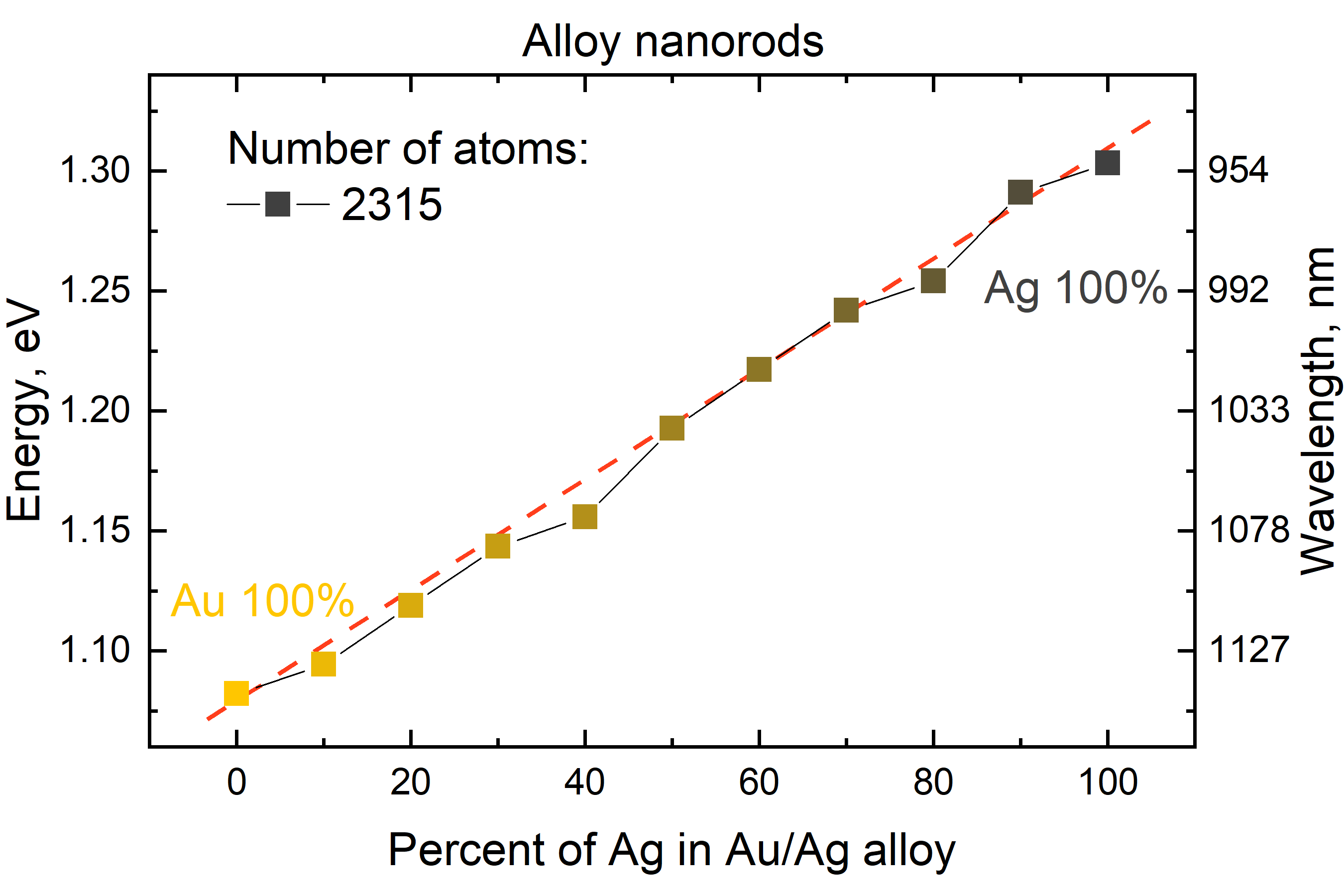}
    \caption{Position of SPR for Au/Ag alloy nanorod with 2315 atoms (squares) with different percentage distribution of Au and Ag along with a dashed line corresponding to Vegard's law.}
    \label{fig:fit_plasmon_rod}
\end{figure}

\begin{figure}
    \centering
    \includegraphics[width = 0.48\textwidth]{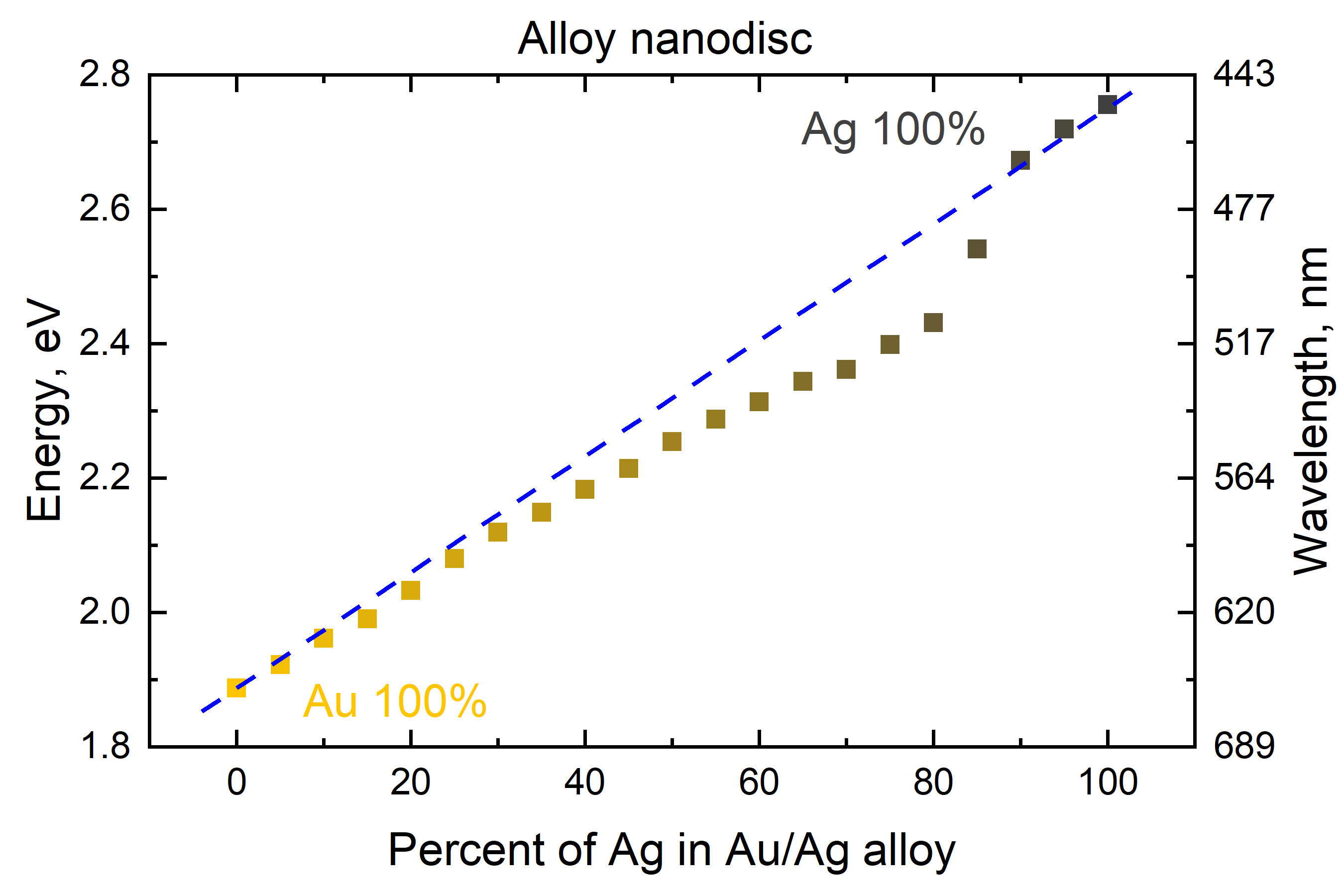}
    \caption{Position of SPR for Au/Ag alloy nanodisc with 4033 atoms with a dashed line showing Vegard's law.}
    \label{fig:sp_alloy_disc}
\end{figure}

From Figs \ref{fig:fit_plasmon_sphere_1553}-\ref{fig:sp_alloy_disc} we show Vegard's law for position of the SPR of alloys with different geometries and distributions. We have in Fig \ref{fig:fit_plasmon_sphere_1553} chosen a random alloy from those sampled in Sec. \ref{sec:stat} and not the average since not all distributions and geometries have been sampled. The same random choice goes for the nanorod and nanodisc alloys in Figs. \ref{fig:fit_plasmon_rod} and \ref{fig:sp_alloy_disc}. One should therefore keep in mind the standard deviation along with the minimum and maximum for the SPR shown in Table \ref{tbl:stat} when interpreting these results.
Comparing the 1553 and 6051 atoms spherical clusters in Fig \ref{fig:fit_plasmon_sphere_1553} we see that changing the size of the sphere does induce any non-linearity nor does the change in geometry which can be seen by comparing all Figs from \ref{fig:fit_plasmon_sphere_1553} to \ref{fig:sp_alloy_disc}. This is in line with other experimental and theoretical works~\cite{Pena-Rodriguez:14,skidanenko,link1999} except that by Nishijima \etal.\cite{Nishijima:12}
In Fig. \ref{fig:sp_alloy_disc} there appears to be some systematic non-linearity in the position of the SPR. By analysing the spectra in Fig. \ref{fig:alloy_disc} in appendix \ref{app:a} we see that the non-linearity comes from the appearance of a shoulder for the pure Ag cluster which turn into a double peak with $15\%$ Au mixed in and finally the shoulder becomes the dominant peak with $20-25\%$ Au in the alloy. This behaviour is not expected to be seen for all discs and the deviation up to 0.2~eV seen is much below the very significant red shift observed by Nishijima \etal.\cite{Nishijima:12}
At the right had side of Figs \ref{fig:fit_plasmon_sphere_1553}-\ref{fig:sp_alloy_disc} the nm scale is also shown. Since the energy range used is small the nm scale will also be almost linear but not completely. This can explain why some have observed weak non-linear trends in the SPR since this will automatically appear if the nm and not the eV energy scale is used.

\begin{figure}
    \centering
    \includegraphics[width = 0.48\textwidth]{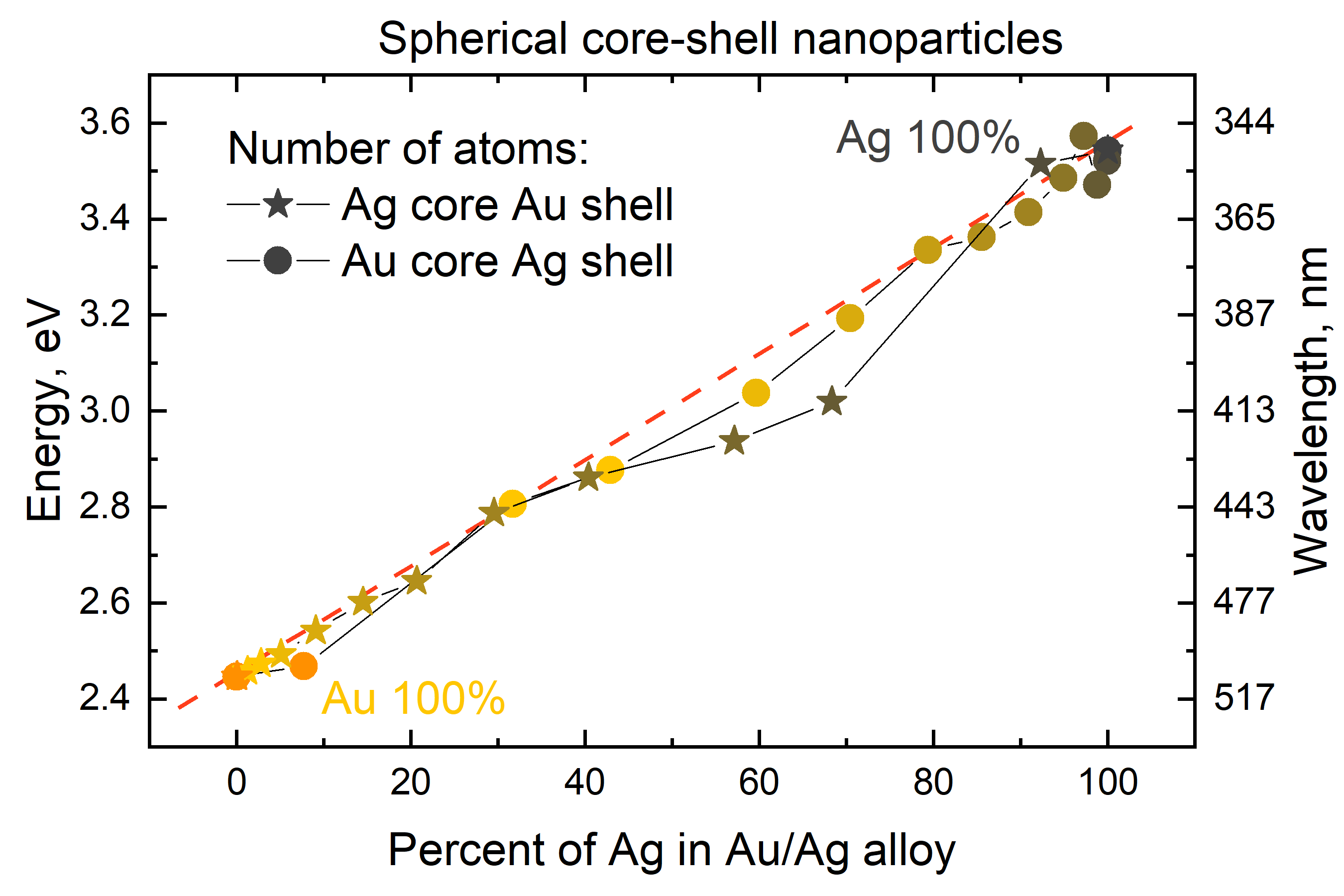}
    \caption{Au-Ag and Ag-Au core-shell nanospheres with different sizes of core and shell as a function the percentage of Ag atoms along with dashed lines showing Vegard's law.}
    \label{fig:sp_core_shell}
\end{figure}

The core-shell structure with an Au core and Ag shell shown in Fig. \ref{fig:sp_core_shell} also follows Vegard's law while the Ag core Au shell core-shell structure shows a small dip of up to 0.2~eV when the number of atoms in the Ag core exceeds 50 and below 80 percent. In this region we, for this cluster, see a large broadening of the spectra as seen in Fig. \ref{fig:curve_ag_core_au_shell} 

\subsection{Polarizability for alloys and core shells}
\label{sec:polarizability}

\begin{figure}
    \centering
    \includegraphics[width = 0.48\textwidth]{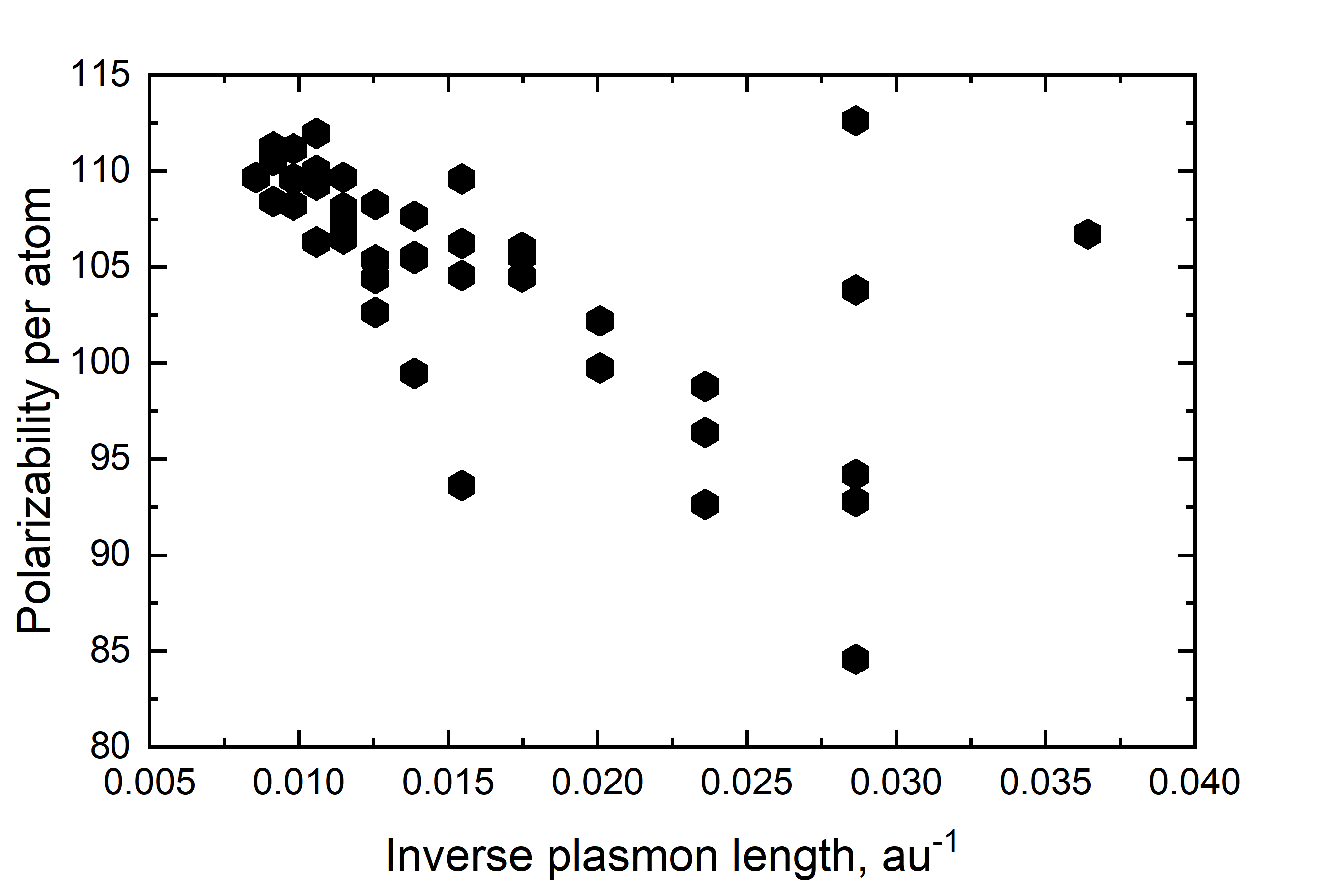}
    \caption{Polarizability per atom for Au as function of inverse plasmon length.}
    \label{fig:pol_devel_inverse}
\end{figure}

For pure metals the polarizability per will decrease proportional to the inverse plasmon length and, in the ex-DIM with the current broadening factor, approach an asymptotic bulk limit of approximately 115~au as seen for Au in Fig. \ref{fig:pol_devel_inverse}. The total polarizability in spheres is therefore in general proportional to the static polarizability and the inverse plasmon length.

Since all alloys and core-shell structures obeyed Vegard's law, except for a small deviation for the nanodisc alloy and the Ag core Au shell as seen in Figs. \ref{fig:sp_alloy_disc} and \ref{fig:sp_core_shell}, and seeing that the polarizability per atom is proportional to the static polarizability it could easily be assumed that the polarizability also would obey Vegard's law    
\begin{equation}
\label{vegard_pol}
p^{Vegard}(x,R) = (1-x)p_{Au}(R) + xp_{Ag}(R),
\end{equation}
where $p$ is the polarizability. This, however, is in general not the case as seen in both experiment and theoretical studies.\cite{Pena-Rodriguez:14,skidanenko,link1999}

\begin{figure}
    \centering
    \includegraphics[width = 0.48\textwidth]{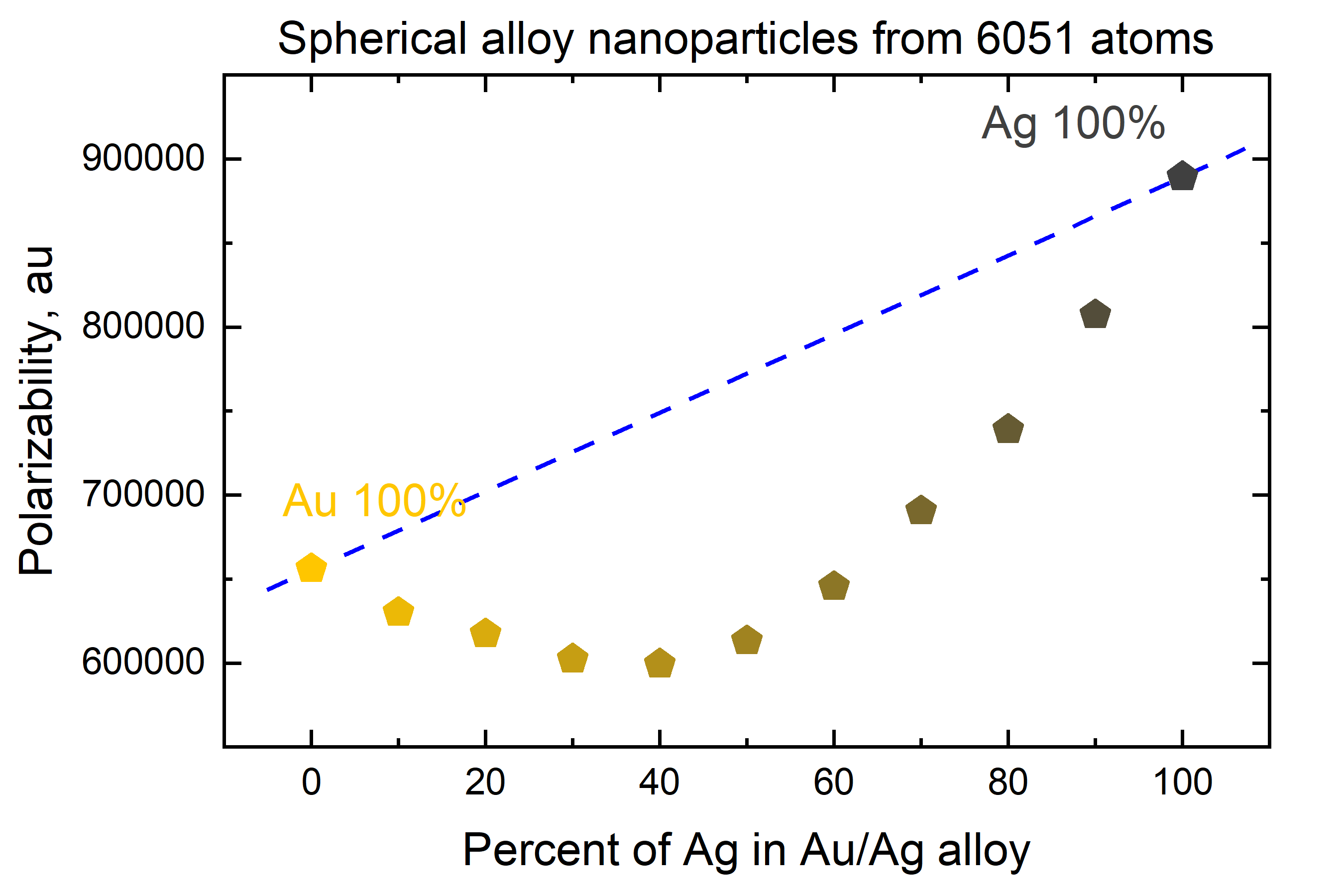}
    \caption{The polarizability at the maximum of the SPR of a 6051 atoms Au/Ag alloy spherical cluster as a function of the percentage of Ag in the cluster along with a dashed line showing Vegard's law.}
    \label{fig:pol_sphere}
\end{figure}

For the Au and Ag sphere alloys a very characteristic dip in the polarizability, as seen in Fig. \ref{fig:pol_sphere}, is observed when adding Ag to an Au cluster despite the fact that Ag has a higher static polarizability. As seen from Fig. \ref{fig:pol_sphere} there is a minimum with 40 percent Ag in the alloy and the polarizability for 30 and 50 percent Ag is very close. The closeness for the polarizability with 30 and 50 percent Ag explains why the extinction cross section per atom in Table \ref{tbl:stat} for these distributions are very close and the 90 percent Ag significantly higher. For the spheres we see exactly the same trend as seen in other studies.\cite{Pena-Rodriguez:14,skidanenko,link1999} 

\begin{figure}
    \centering
    \includegraphics[width = 0.48\textwidth]{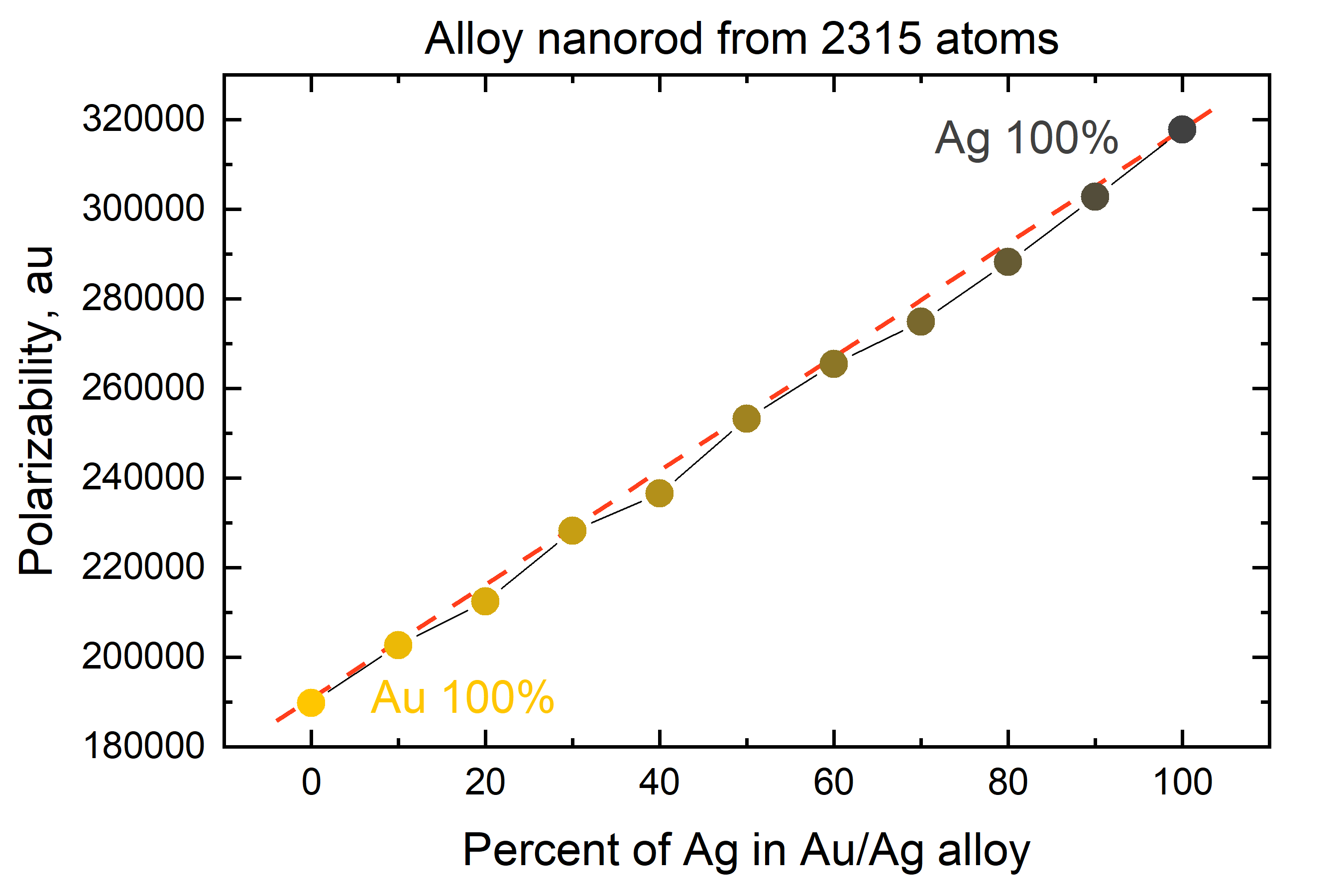}
    \caption{The maximum value of the polarizability as a function of percent of Ag in a 2315 atoms Au/Ag alloy nanorod with an aspect ratio $5.4$.}
    \label{fig:pol_rod}
\end{figure}

For the rod alloy, seen in Fig. \ref{fig:pol_rod}, we observe a perfect linear correlation of the polarizability with the distribution for the longitudinal SPR. The longitudinal SPR of the rod alloy is the only alloy geometry, that we have tried, where the polarizability follows Vegard's law and the only geometry where there is no loss in response to the external field in comparison to Vegard's law.

\begin{figure}
    \centering
    \includegraphics[width = 0.48\textwidth]{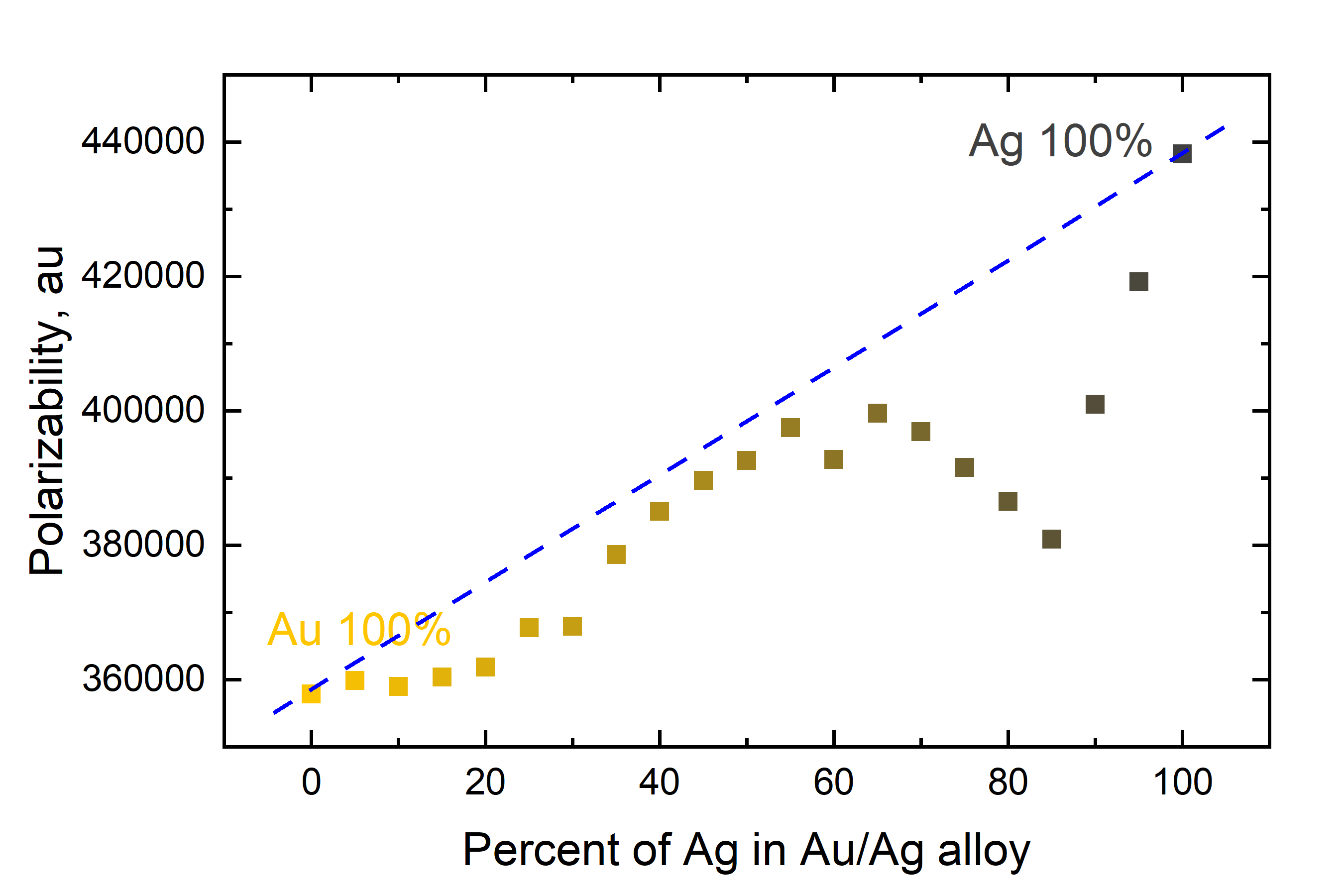}
    \caption{The maximum polarizability as a function of the percentage of Ag in Au/Ag alloy nanodisc with 4033 atoms with a dashed line showing Vegard's law.}
    \label{fig:pol_alloy_disc}
\end{figure}

The nanodisc alloy, seen in Fig. \ref{fig:pol_alloy_disc}, shows a very large dip in the maximum polarizability when having 10-30 percent Au in the alloy. The dip in the maximum polarizability is again related to the appearance of a shoulder and double peak as seen in Fig. \ref{fig:alloy_disc} where the peak is significantly broader than for the rest of the nanodisc alloys. 

\begin{figure}
    \centering
    \includegraphics[width = 0.48\textwidth]{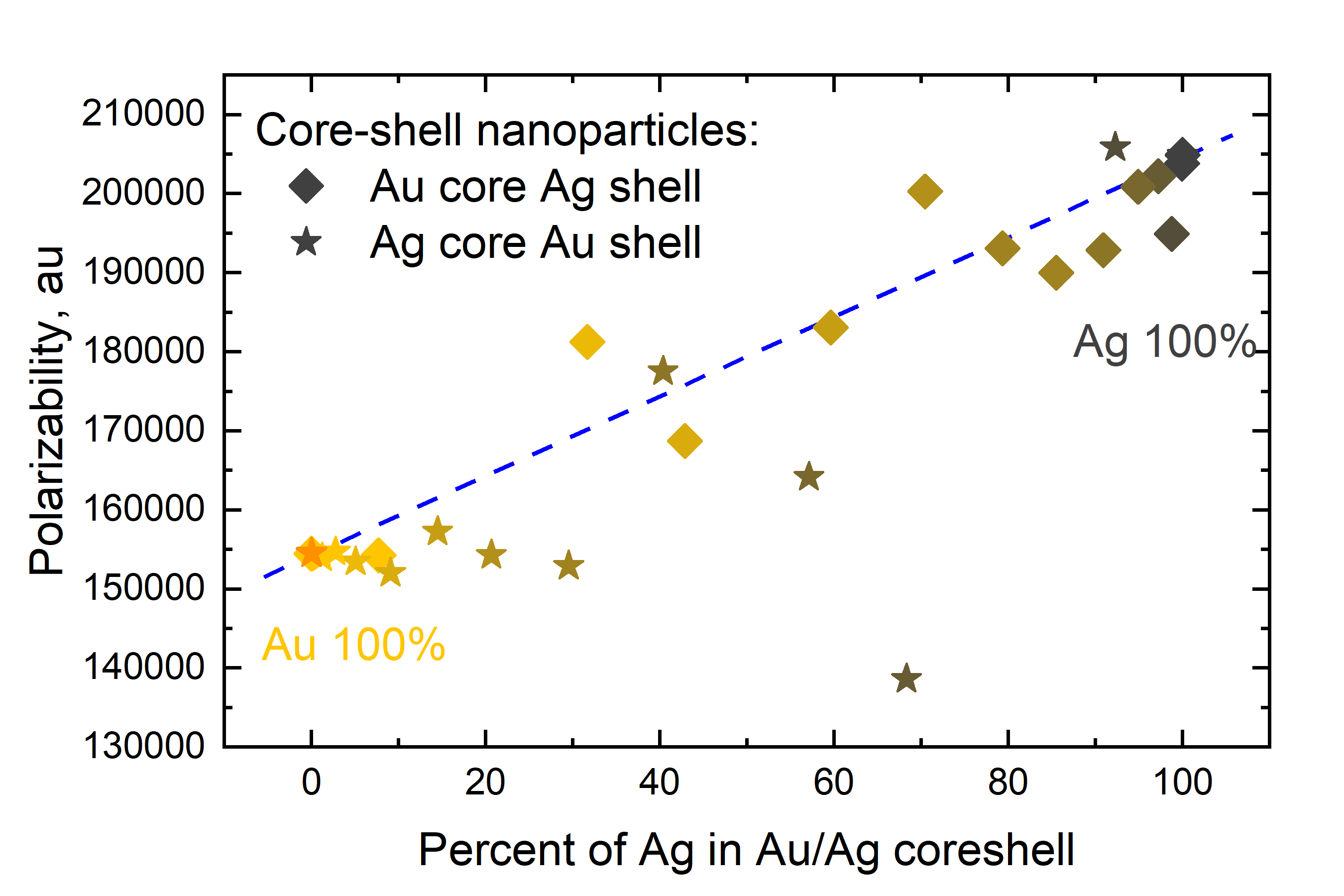}
    \caption{The maximum polarizability as a function of the percentage of Ag/Au and Au/Ag spherical cluster with 1553 atoms along with a dashed line showing Vegard's law.}
    \label{fig:pol_au_core_ag_shell}
\end{figure}

For the core-shell structures shown in Fig. \ref{fig:pol_au_core_ag_shell} there is no obvious trend and larger variations from small changes in the distribution is seen. The large dip is seen for the large Ag core and a thin Au shell where the number of atoms in the core is just below 70 percent. This is again caused by a large broadening of the spectra as seen in Fig. \ref{fig:curve_ag_core_au_shell}. As the only one that we have found does the Au core Ag shell nanoparticle have a few sizes of core and shell where the polarizability is above that predicted by Vegard's law though nothing systematic.

\section{Conclusion}

We here present a way of creating accurate calculations on alloys using a discrete atomic structure model with good estimation of error bars. This was achieved by using a simple random number generator and distribution for the constituents in our extended discrete interaction model (ex-DIM) in the initial placement of atoms in the alloy. We lay out the physical reasoning of why it is appropriate to use a perfect lattice for all sizes of both pure metals and alloys which is backed up by a statistical analysis of the alloys from where error bars can be estimated from. We see the trend in the error follows that expected trend from the fitting of the size dependence of Au clusters and statistics where the standard deviation of the surface plasmon resonance (SPR) increases inversely with size and evenness of the constituents in the alloy. 

Much of the motivation for creating alloys or core-shell structures is motivated by the ability of also blue shifting the SPR and having a chemically less reactive surface while still having significant response to the impinging light. Another motivating fact was the experimental report by Nishijima \etal \cite{Nishijima:12} on the breaking of Vegard's law\cite{vegard,vegards_law} for the SPR simply from geometrical alterations. We have here examined three Au-Ag alloys with different geometrical structures namely spheres, a rod and  a disc and have found no deviations from Vegard's law for the SPR, using a linear energy unit, that is not readily explained by the estimated error bars or from the emergence of a double peak. For spheres this is in line with other experimental and theoretical predictions.~\cite{Pena-Rodriguez:14,skidanenko,link1999} 
For the discs we were not able to reproduce the experimental findings by Nishijima \etal \cite{Nishijima:12} even if the atomic interaction in the ex-DIM should be able to simulate the polarization of the individual atoms. The idealized structure of alternating Ag and Au atoms giving a closed shell pair structure of Ag and Au envisioned by Nishijima \etal as an explanation of the large red shift of the SPR will of course never be seen by a random distribution and we have not observed any extremely red shifted outliers in our data.
For spherical core-shell structures the Au core Ag shell also shows agreement with Vegard's law while the Ag core Au shell shows a red shift of up to 0.2~eV compared to the expected from Vegard's law for large Ag cores though these peaks showed much larger broadening and these was no systematic behaviour in the evolution of the SPR with variations of the size of the core could be observed.

While the SPR, in all cases tested here, closely follow Vegard's law this was not the case for the maximum value of the polarizability which shows a great dependence on the geometry of the nanoparticle. For the alloys there in general was a slight broadening of the spectra leading to a lower maximum value of the polarizability. This showed up systematically for the spherical alloys where there was a minimum in the polarizability at 40 percent Ag in the Au/Ag alloy irrespective of size. The nanorod structure was the only alloy which followed Vegard's law and the polarizability was not below that predicted by Vegard's law. For the nanodisc we, in this case, saw a sudden dip in the polarizability due to having a double peak for alloys containing 75-90 percent Ag. The polarizability for the core-shell structure showed a very unsystematic nature and it would therefore be hard to predict the polarizability for these.   

From an atomistic perspective we see that the classical way of treating an alloy without any resolution at the atomic level does not introduce any significant error for the SPR of larger systems calculated using classical methods.\cite{silver_dielectric_2014,doi:10.1021/acs.jpcc.0c02630,Kuladeep:s} Provided that the size correction for the dielectric function is good then there should be no problems for classical methods in simulating nanoparticle alloys down to the 4-5nm or even smaller depending on the distribution of elements since at these sizes we still see very small standard deviation in our calculations.

\section*{Conflicts of interest}

There are no conflicts of interest to declare.

\section*{Acknowledgements}

H.\r A. and V.Z. acknowledge the support of the Russian Science Foundation (project No. 18-13-00363). 
L.K.S acknowledges the support of Carl Tryggers Stifetelse, project CTS 18-441.
The simulations were performed on resources provided by the Swedish National Infrastructure for Computing (SNIC) at NSC under the project 
 "Multiphysics Modeling of Molecular Materials",  SNIC 2019/2-41.

\section*{Appendix A: Complete curves for alloys and core-shell nanoparticles}
\label{app:a}

\begin{figure}
    \centering
    \includegraphics[width = 0.48\textwidth]{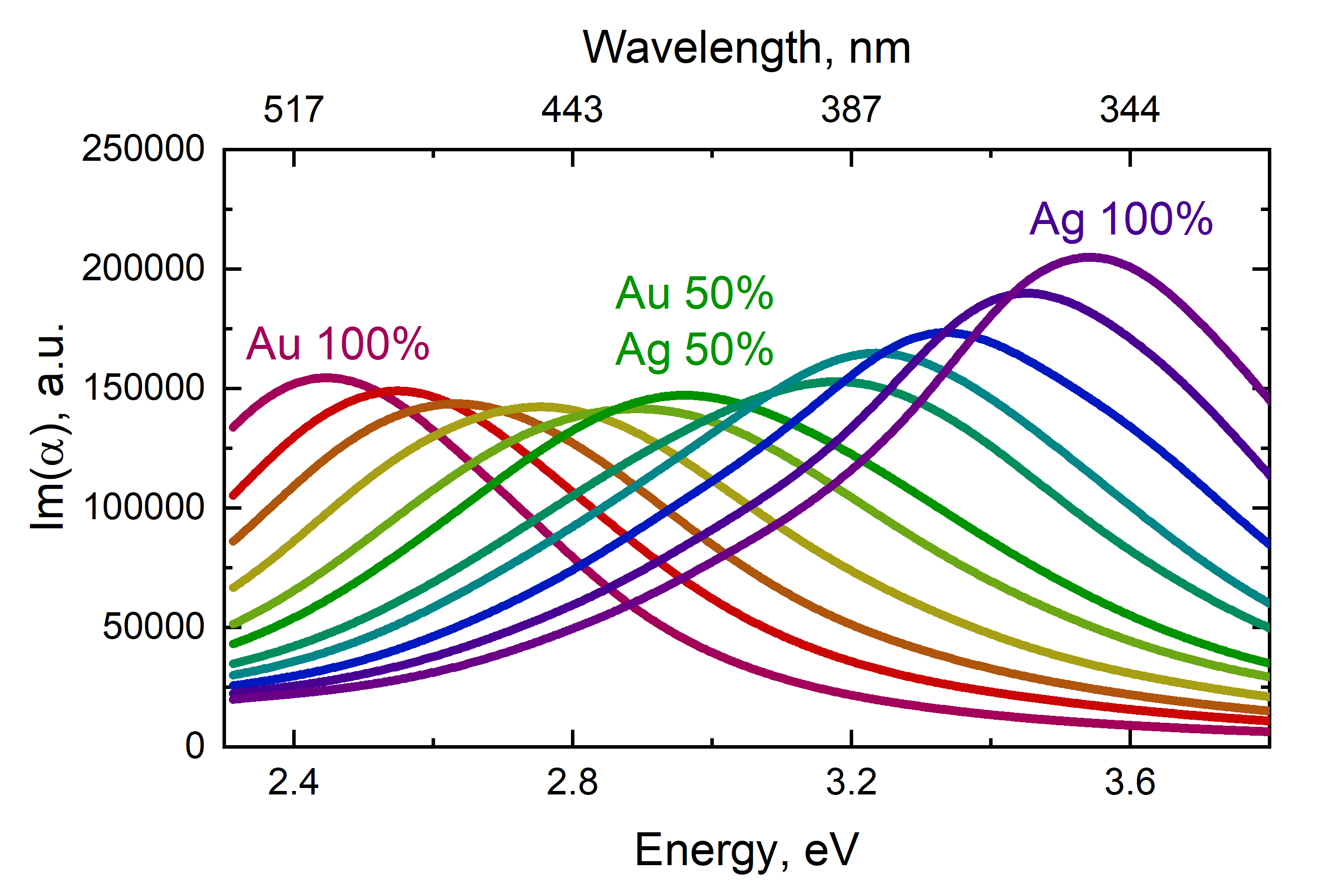}
    \includegraphics[width = 0.48\textwidth]{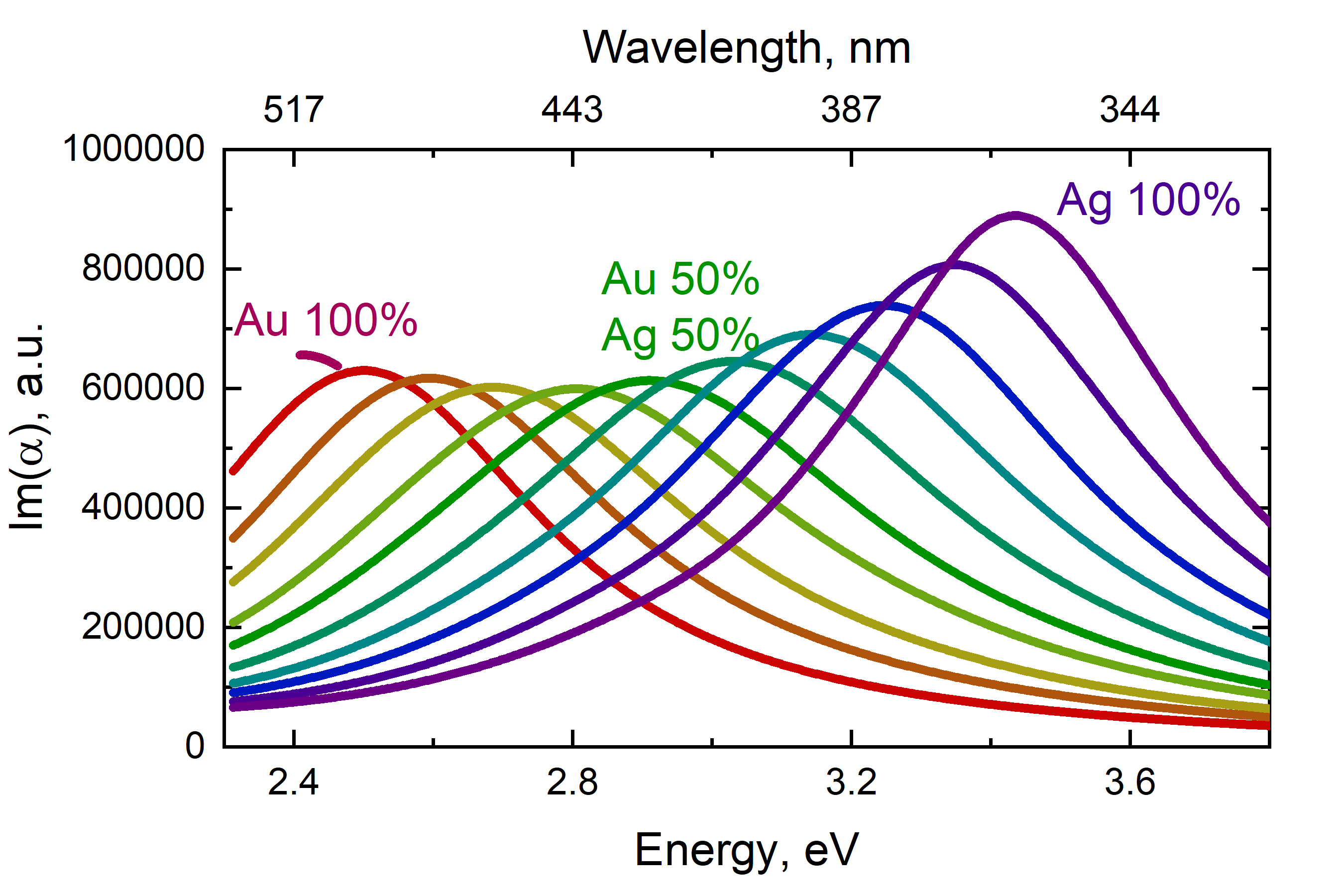}
    \caption{Spherical alloy Au-Ag nanoparticle from 1553 (top) and 6051 (bottom) atoms with different distributions.}
    \label{fig:spec_alloy}
\end{figure}

We here show the complete curves for all alloys and core-shell nanoparticles. For the spherical alloys seen in Fig. \ref{fig:spec_alloy} we see that the shape of the spectra is not altered by the mixing of Au and Ag and Vegard's law for the SPR is obeyed. The dip in the polarizability as discussed in Sec. \ref{sec:polarizability} and plotted in Fig. \ref{fig:pol_sphere} for both spheres is also visible in Fig \ref{fig:spec_alloy}. The fact that the width does change can be caused by the fact that the broadening factor used for Au and Ag is the same.  

\begin{figure}
    \centering
    \includegraphics[width = 0.48\textwidth]{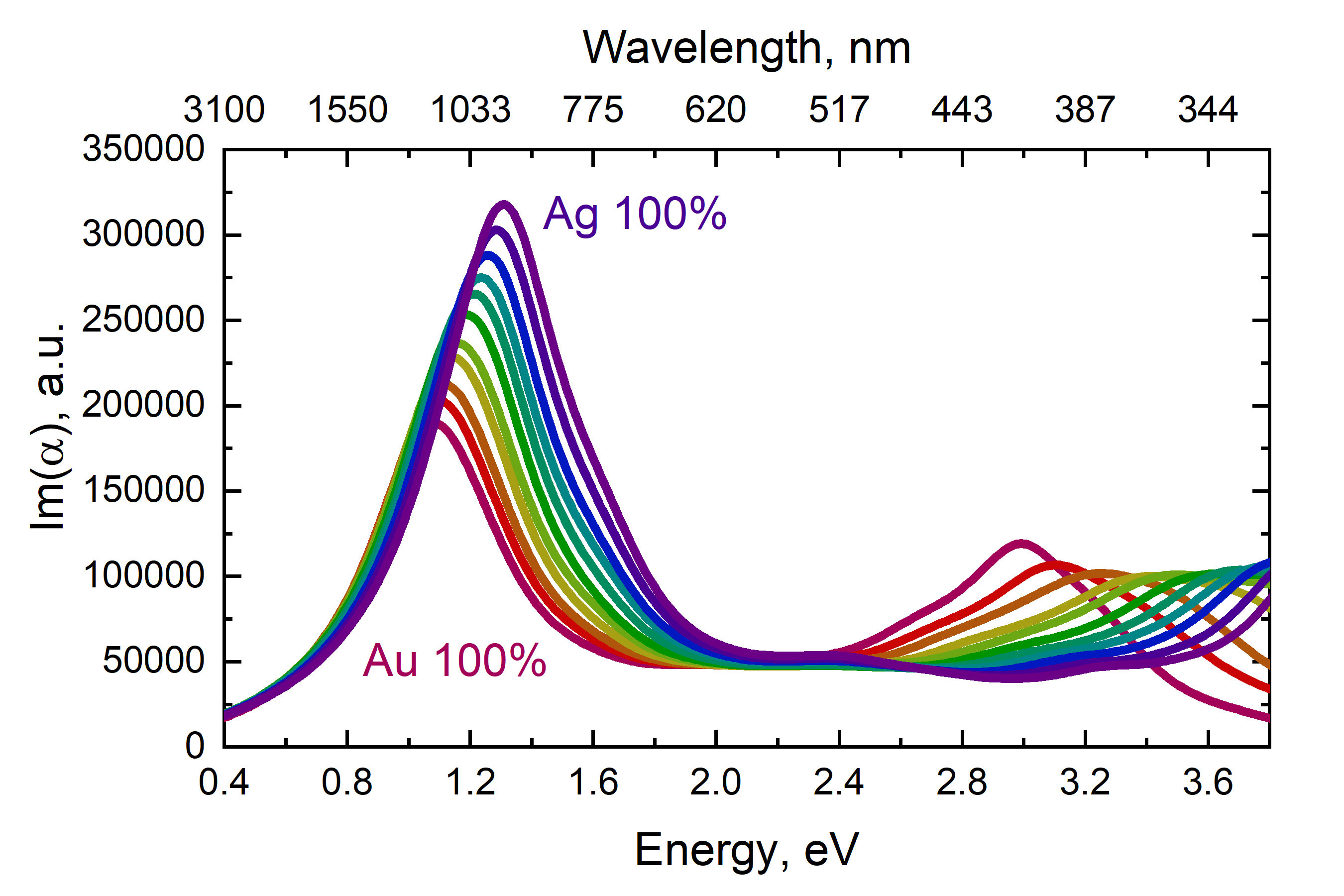}
    \caption{Optical spectra for different distributions of Au/Ag nanorod alloy with 2315 atoms and aspect ratio of 5.4.}
    \label{fig:alloy_curve_nm_rod}
\end{figure}

For the nanorod alloy in Fig. \ref{fig:alloy_curve_nm_rod} both the longitudinal and transverse SPR are clearly visible. From Fig. \ref{fig:alloy_curve_nm_rod} it is seen that Vegard's law is obeyed for both the SPR and polarizability for the longitudinal SPR. For the transverse only the SPR follows Vegard's law. The nanorod alloy is the only nanoparticle, we have found, where Vegard's law is obeyed for both the SPR and the polarizability and also the only nanoparticle alloy where the polarizability is not below Vegard's law. 

\begin{figure}
    \centering
    \includegraphics[width = 0.48\textwidth]{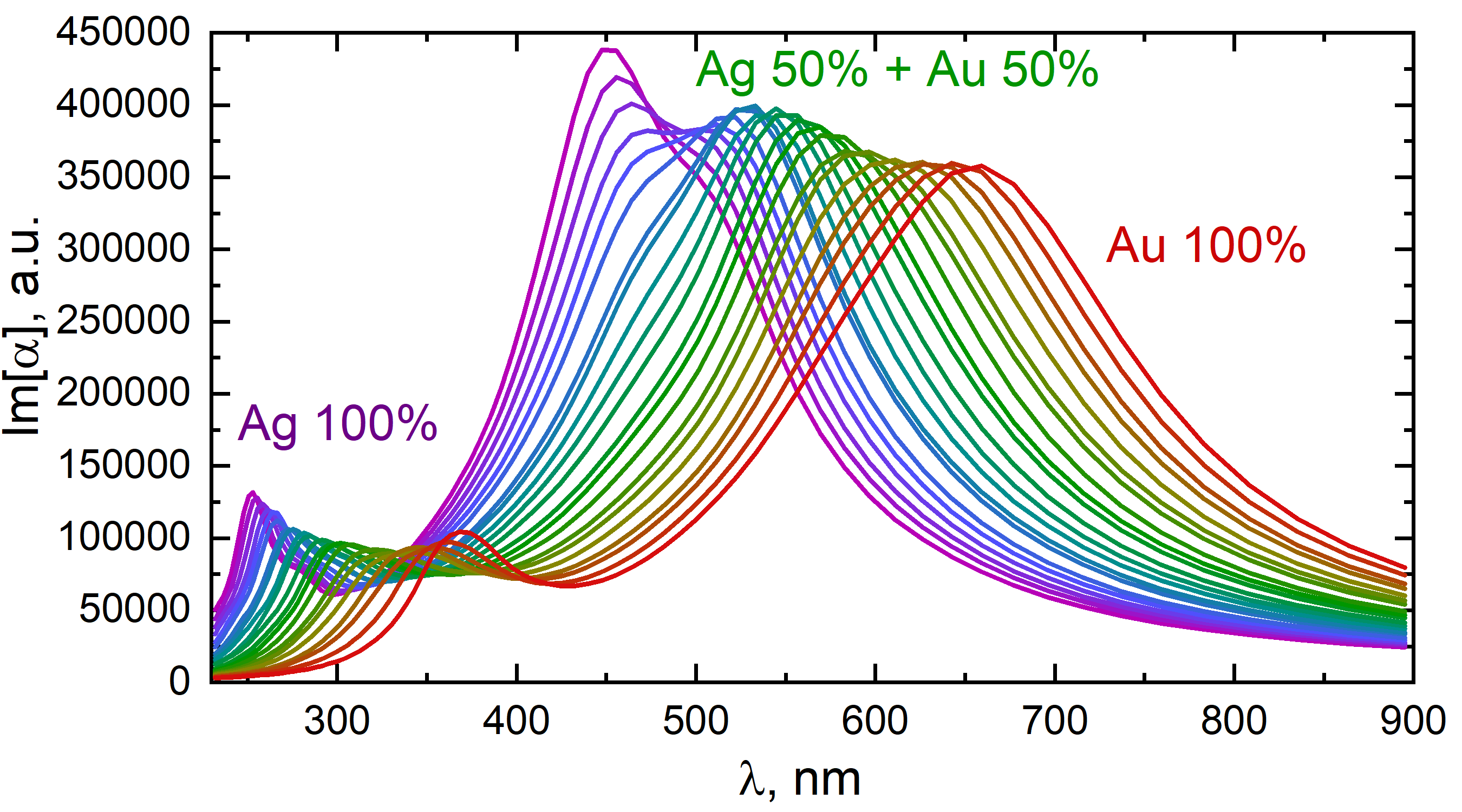}
    \caption{Alloy disc from 4033 atoms with diameter = 9nm, and height = 4nm. }
    \label{fig:alloy_disc}
\end{figure}

The longitudinal and transverse SPR is also visible for the nanodisc alloy in Fig. \ref{fig:alloy_disc}. We her clearly see what appear to be some systematic deviation from Vegard's law in Fig. \ref{fig:sp_alloy_disc} is due to the appearance of a shoulder for the pure Ag cluster which turn into a double peak with $15\%$ Au mixed in and finally the shoulder becomes the dominant peak with $20-25\%$ Au in the alloy. The shoulder and double peak also explains the sharp drop in the maximum value for the polarizability in Fig. \ref{fig:pol_alloy_disc}. So while a small and systematic deviation from Vegard's law is observed here the deviation is far from that observed by Nishijima \etal where the red shift in the $50\%$ of both Au and Ag is below the pure Au peak.\cite{Nishijima:12}

\begin{figure}
    \centering
    \includegraphics[width = 0.48\textwidth]{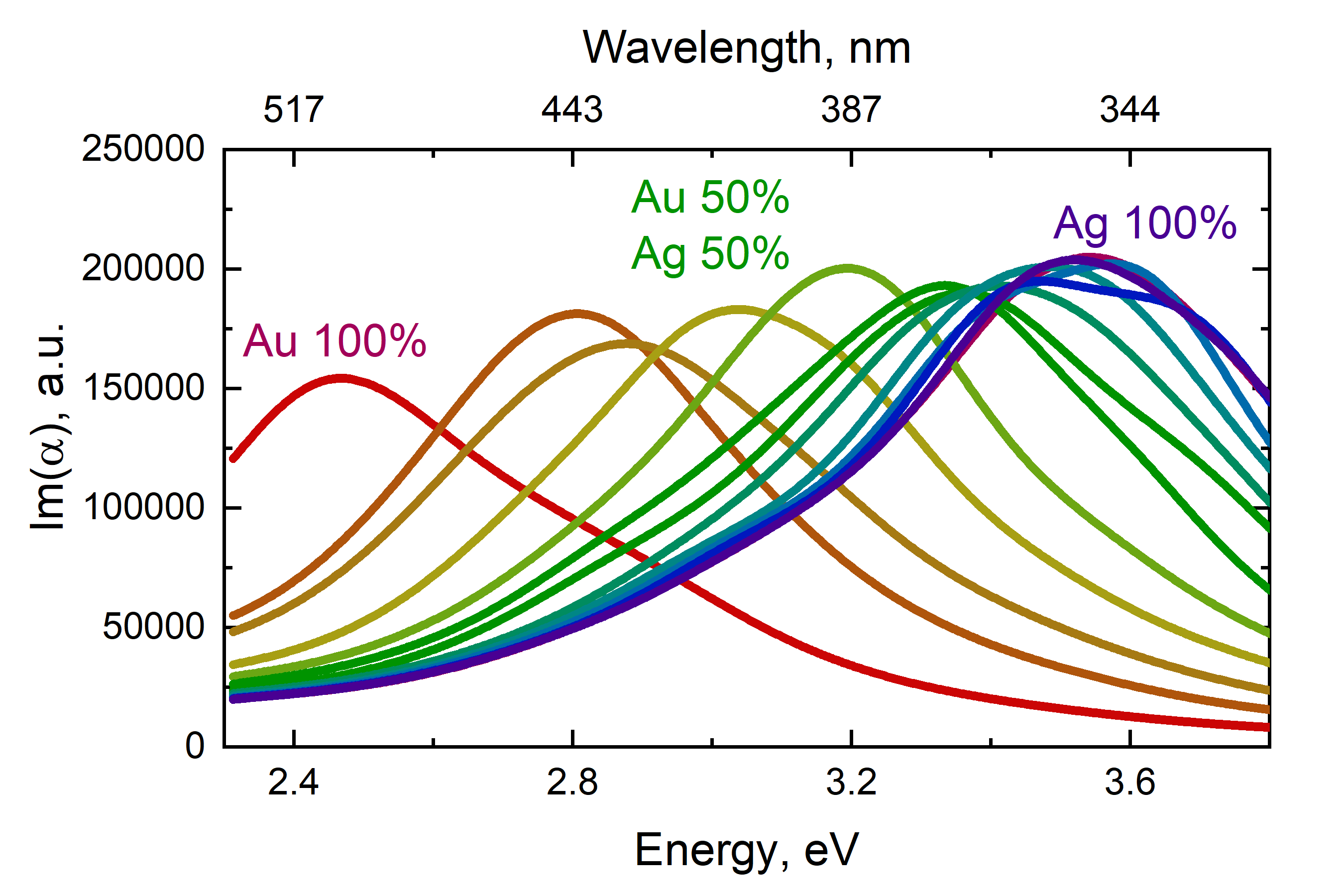}

    \includegraphics[width = 0.48\textwidth]{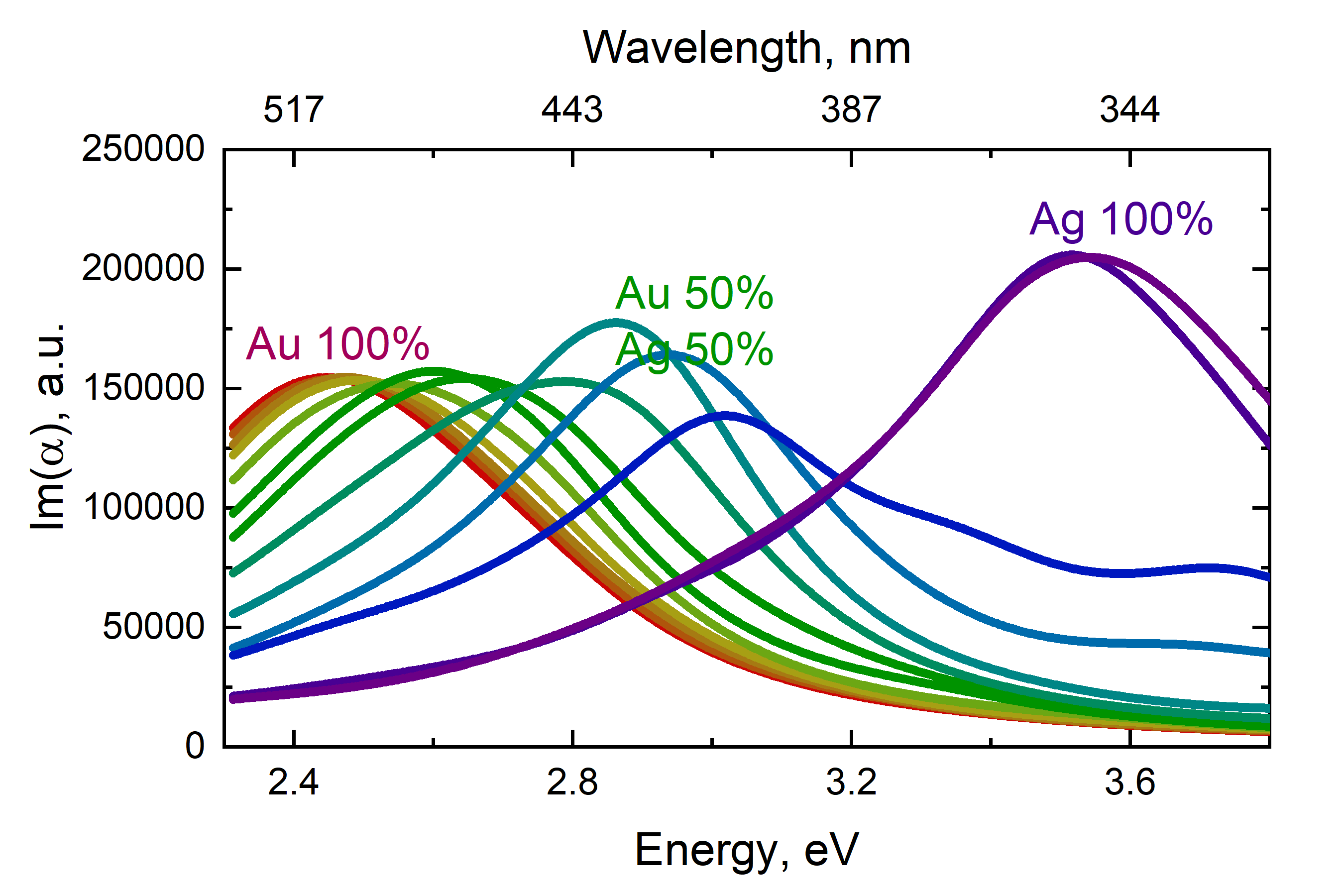}
    \caption{Optical spectra of Au core and Ag shell (top) and Ag core and Au shell (bottom) spherical nanoparticles with 1553 atoms.}
    \label{fig:curve_ag_core_au_shell}
\end{figure}

For the core-shell spectra in Fig. \ref{fig:curve_ag_core_au_shell} no real pattern emerges in the progression from pure Au to pure Ag clusters since both the maximum polarizability and the FWHM changes rapidly. Due to the atomistic nature of the ex-DIM having smaller steps in some of the areas where there is a rapid change can be difficult.

\balance

\bibliography{rsc}
\bibliographystyle{rsc}

\end{document}